\documentclass[oneside,reqno]{amsart}
\usepackage{amsmath,amsthm,amssymb,amsfonts,mathrsfs,enumerate,graphicx}
\newcommand{\dummy}{\protect \rule{0.0em}{1.0ex}}
\newcommand{\Dummy}{\protect \rule{0.0em}{1.6ex}}

\newcommand{\pretwo}{\,\Dummy^{(2)}}
\newcommand{\prethree}{\,\Dummy^{(3)}}

\def \smallfrac#1#2{{\textstyle \frac{#1}{#2}}}  
\protect
\protect\newtheorem*{corollary*}{Corollary}
\protect
\protect\newtheorem*{sublemma*}{Sub-lemma}
\protect\newtheorem{theorem}{Theorem{}\protect\setcounter{corollary}{0}\setcounter{sublemma}{0}}[section]
\protect\newtheorem*{theorem*}{Theorem{}}
\protect\newtheorem{lemma}[theorem]{\setcounter{corollary}{0}Lemma}
\protect\newtheorem*{lemma*}{\setcounter{corollary}{0}Lemma}
\protect\newtheorem{proposition}[theorem]{\setcounter{corollary}{0}Proposition}
\protect\newtheorem*{proposition*}{\setcounter{corollary}{0}Proposition}
\protect\newtheorem{definition}[theorem]{Definition}
\protect\newtheorem*{definition*}{Definition}
{\protect\theoremstyle{remark}}
{\protect\theoremstyle{remark}\newtheorem*{scholium*}{Scholium}}
{\protect}
{\protect\newtheorem{conjecture*}[theorem]{\bf Conjecture}}
\protect\newtheorem*{remark*}{Remark}
{\theoremstyle{remark}}
\renewenvironment{proof}[1][]{\noindent{\em Proof#1. }}{\mbox{}\hspace{\fill}\qedsymbol \\[-0.5ex] \mbox{}}

\newcommand{\saveqedsymbol}{\qedsymbol}
\newcommand{\subqedsymbol}{\raisebox{0.5ex}{$\bigtriangledown$}}
{\begin{proof}[#1]}%
{\renewcommand{\qedsymbol}{\subqedsymbol}\end{proof}\renewcommand{\qedsymbol}{\saveqedsymbol}}

\newcommand{\noproof}{\hspace{\fill}\qedsymbol \vspace{1ex}}
\theoremstyle{remark}
\theoremstyle{remark}
\theoremstyle{remark}\newtheorem{example}{{\bf Example}}
\begin{document}
\pagenumbering{arabic}
\title[Photon surfaces]{}
\begin{center}
{\Large \bf THE GEOMETRY OF PHOTON SURFACES}\\[6ex]
{CLARISSA-MARIE CLAUDEL\footnote{\noindent e-mail: {\tt clarissa@abel.math.umu.se}}}\\
{\small \em Mathematical Institute, University of Ume{\aa}, S-901 87 Ume{\aa}, Sweden}\\[2ex]
{K. S. VIRBHADRA\footnote{e-mail: {\tt shwetket@maths.uct.ac.za}}
and G. F. R. ELLIS\footnote{e-mail: {\tt ellis@maths.uct.ac.za}}\\
{\small \em Department of Applied Mathematics, University of Cape
Town,\\ Rondebosch 7701, South Africa.}}  \vspace{-5ex}
\end{center}
\begin{abstract}
The photon sphere concept in Schwarzschild space-time is
generalized to a definition of a photon surface in an arbitrary
space-time. A photon sphere is then defined as an $SO(3)\times 
{\mathbb{R}}$-invariant photon surface in a static spherically
symmetric space-time.   It is proved, subject to an energy
condition, that a black hole in any such space-time must be
surrounded by a photon sphere. Conversely, subject to an energy
condition, any photon sphere must surround a black hole, a naked
singularity or more than a certain amount of matter. A second
order evolution equation is obtained for the area of an
$SO(3)$-invariant photon surface in a general non-static
spherically symmetric space-time. Many examples are provided.
\end{abstract}
\maketitle
\protect\section{Introduction} The exterior region of the
maximally extended Schwarzschild space-time is described by the
metric
  \begin{equation} {\bf
 g} = -\left(1-\frac{2m}{r}\right) dt^2  + \left(1-\frac{2m}{r}\right)^{-1}
dr^2
       + r^2 \left(d\theta^2 + \sin^2\theta d\phi ^2 \right),  \quad
\quad    r > 2m \ .
  \label{sch}
 \end{equation}
For any null geodesic in this exterior region the null geodesic
equations give
\begin{equation}
\frac{d^2r}{d\lambda^2} = \left(r-3m\right)
   \left\{
         \left(\frac{d \theta}{d \lambda}\right)^2
         + \sin^2\theta \left(\frac{d \phi}{d \lambda}\right)^2
    \right\}
\end{equation}
where $\lambda$ is an affine parameter along the geodesic. The
right side here is evidently positive for $r > 3m$ and negative
for $r : 2m <r < 3m$. It follows that any future endless null
geodesic in the maximally extended Schwarzschild space-time
starting at some point with $r > 3m$ and initially directed
outwards, in the sense that $dr/d\lambda$ is initially positive,
will continue outwards and escape to infinity. Any future endless
null geodesic in the maximally extended Schwarzschild space-time
starting at some point with $r: 2m <r <3m$ and initially directed
inwards, in the sense that $dr/d\lambda$ is initially negative,
will continue inwards and fall into the black hole. The
hypersurface $\{r = 3m\}$, known as the Schwarzschild photon
sphere, thus distinguishes the borderline between these two types
of behaviour; any null geodesic starting at some point of the
photon sphere and initially tangent to the photon sphere will
remain in the photon sphere. (See Darwin \protect\cite{Dar1,Dar2}
for a detailed analysis of the behaviour of null and timelike
geodesics in Schwarzschild space-time.)

The Schwarzschild photon sphere also has physical significance for
massive bodies. For any timelike geodesic in the exterior region
the geodesic equations give
\begin{equation}
\frac{d^2r}{ds^2} = -\frac{m}{r^2} + \left(r-3m\right)
   \left\{
         \left(\frac{d \theta}{ds}\right)^2
         + \sin^2\theta \left(\frac{d \phi}{ds}\right)^2
    \right\}
\label{timegeodesic}\end{equation} where $s$ is arc length along
the geodesic.  At any point with $r>3m$ one may arrange for the
two terms on the right of (\protect\ref{timegeodesic}) to cancel
and so obtain a timelike geodesic at constant $r$.  For $r:
2m<r<3m$ the right hand side of (\protect\ref{timegeodesic}) is
evidently negative. Thus any future endless timelike geodesic in
the maximally extended Schwarzschild space-time starting at some
point between the event horizon at $r = 2m$ and the photon sphere
at $r = 3m$ and initially directed inwards, in the sense that
$dr/ds$ is initially negative, will continue inwards and fall
into the  black hole. Any observer who traverses a Schwarzschild
photon sphere must therefore engage some form of propulsion or
else be drawn in to the black hole to meet an inevitable fate.

A photon sphere has been defined by Virbhadra \& Ellis
\protect\cite{VE99a} as a timelike hypersurface of the form $\{ r
= r_0 \}$ where $r_0$ is the closest distance of approach for
which the Einstein bending angle of a light ray is unboundedly
large.  These authors subsequently \protect\cite{VE99b}
considered the Einstein deflection angle for a general static
spherically symmetric metric and obtained an equation for a
photon sphere. The existence of a photon sphere in a space-time
has important implications for gravitational lensing. In any
space-time containing a photon sphere, gravitational lensing will
give rise to relativistic images \protect\cite{VE99a}.

The Schwarzschild photon sphere may be usefully be compared with
the concept of a closed trapped surface.  Any null geodesic
originating from any point on a closed trapped surface in
Schwarzschild space-time is drawn into the singularity at $r=0$.
By contrast, any null geodesic originating from any point on
the photon sphere will be drawn into the singularity if and only
if it is initially directed inwards.

The main objectives of the present paper are to give a geometric
definition of a photon surface in a general space-time and of a
photon sphere in a general static spherically symmetric
space-time.

An evolution equation is obtained for the cross-sectional area of
a photon surface in a dynamic spherically symmetric space-time.
It is shown, subject to suitable energy conditions, that in any
static spherically symmetric space-time a black hole must be
surrounded by a photon sphere, and a photon sphere must surround
either a black hole, a naked singularity or more than a certain
amount of matter.  Many examples are given of photon spheres in
static spherically symmetric space-times. Photon surface evolution
is considered for the dynamic space-time example of Vaidya null
dust collapse to a naked singularity.
\protect\section{Photon surfaces}\label{photonsurfacesection} The
hypersurface $S:=\{ r=3m\}$ in Schwarzschild space-time has two
main properties, first that any null geodesic initially tangent
$S$ will remain tangent to $S$, and second that $S$ does not
evolve with time. The following general definition of a photon
surface is based on only the first of these properties.  A more
restrictive class of photon surfaces may be defined when the
space-time admits a group of symmetries (see Definition
\protect\ref{photonspheredef}).
\begin{definition} \label{photonsurfacedef} {\rm A photon surface
of $(M,{\bf g})$ is an immersed, nowhere-spacelike hypersurface
$S$ of $(M, {\bf g})$ such that, for every point $p \in  S$ and
every null vector ${\bf k} \in T_pS$, there exists a null geodesic
$\gamma : (-\epsilon, \epsilon ) \rightarrow M$ of $(M,{\bf g})$
such that $\dot{\gamma}(0) = ${\bf k}$, |\gamma| \subset S$.}
\end{definition}

Any null hypersurface is trivially a photon surface. Photon
surfaces are conformally invariant structures. If $S$ is a photon
surface of $(M,{\bf g})$ then $S$ is a photon surface of
$(M,\Omega^2 {\bf g})$ for any smooth function $\Omega : M
\rightarrow (0,\infty )$.

Note that Definition \protect\ref{photonsurfacedef} is entirely
local. In particular, a photon surface $S$ need contain no
endless null geodesics of $(M,{\bf g})$. Moreover, a photon
surface need only be immersed, rather than embedded in $M$, and
so may have self-intersections. If $(M,{\bf g})$ is of dimension
$n+1$ ($n \geq 2$) then, through each point $p$ of a photon
surface $S$ in $(M,{\bf g})$, there is an $(n-2)$-parameter
family of null geodesics of $(M,{\bf g})$ that lie entirely in
$S$.

The paper will be principally concerned with photon surfaces in
space-times of $3+1$ dimensions.  The exceptions are Examples
\protect\ref{Min3} and \protect\ref{deSitter} which give photon
surfaces in space-times of dimension $2+1$ and $4+1$ respectively.

\begin{example}[Minkowski 3-space] \label{Min3}
{\rm In Minkowski 3-space ${\mathbb{M}}^3$, consider the
single-sheeted hyperboloid $S$ given by
\begin{equation}
-t^2 + x^2 + y^2 = a^2 \label{eq2}
\end{equation}
for some constant $a > 0$. This surface is doubly ruled, the
rulings being given by
\begin{equation}
\gamma^{\pm}_{\theta} \left(t \right) := a \left(0,\cos \theta,\sin
\theta \right)
  +at \left(1,\mp \sin \theta, {\pm}\cos \theta \right)
\label{eq3}
\end{equation}
($-\infty < t < \infty, 0 \leq \theta < 2 \pi$), where $\theta$
identifies the intersection points with $ \{t =0\} $ and $t$ is
the parameter along the ruling lines. The tangents
$\dot{\gamma}^{\pm}_{\theta}\left(t \right)$ to the ruling lines  are
null  with respect to the ${\mathbb{M}}^3$ metric. Clearly they
are geodesics in ${\mathbb{M}}^3$. At each point of $S$ there can
be just two null directions tangent to $S$. These must therefore
be the directions of the two ruling lines through that point.
Hence $S$ is a photon surface in the sense of Definition
\protect\ref{photonsurfacedef} (see Fig.~\protect\ref{claire1}).

\begin{figure}[t]
\begin{center}
\includegraphics[width=3.0in]{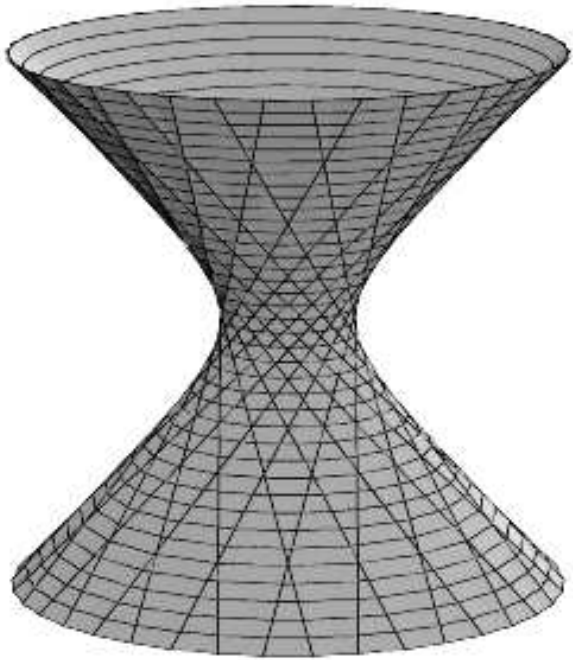}
\end{center}
\caption{A photon surface in Minkowski $3$-space. The
two families of ruling lines are null geodesics with respect to
both the Minkowski $3$-metric and the induced $2$-metric.  The
lines may be regarded as the space-time paths of pulsed laser
beams.}
\label{claire1}
\end{figure}

Note that for any circle of the form
\begin{equation}
C = \left\{ t_0, x_0+r \cos \theta, y_0+r \sin \theta \right\},
\quad \quad  r > 0, \label{eq4}
\end{equation}
and any future-directed timelike vector field ${\bf X}$ along $C$
that respects the symmetry of $C$, in the sense of
\begin{equation}
{\bf X} = \left(X^t, X^r \cos \theta, X^r \sin \theta \right),
\label{eq5}
\end{equation}
for constant $X^t>0$, $X^r$ such that $(X^t)^2 >(X^r)^2$, there
is a unique single-sheeted hyperboloid $S$ through $C$ such that
${\bf X}$ is tangent to $S$ along $C$.

In the case $a=0$, equation (\protect\ref{eq2}) gives the null
cone through the origin.  The complement of $p$ in this null cone
is a null photon surface of ${\mathbb{M}}^3$.  }
\end{example}

\begin{example}[Minkowski 4-space]\label{Min4} {\rm One may generalize
Example 1 to Minkowski 4-space ${\mathbb{M}}^4$ as follows. Let
$S$ be a timelike hypersurface in ${\mathbb{M}}^4$ of the form
\begin{equation}
-t^2+x^2+y^2+z^2 = a^2 \label{ellhyp4}
\end{equation}
for some constant $a>0$. The two-parameter family of lines
\begin{equation}
\gamma^\pm_{\theta,\phi}\left(t \right) =
     a \left( 0,\cos \theta , \sin \theta \sin \phi , \sin \theta \cos \phi \right)
 +at \left( 1,\mp \sin \theta , {\pm}\cos \theta \sin \phi ,{\pm}\cos \theta \cos \phi \right)
\label{eq7}
\end{equation}
foliate $S$ and are null geodesics with respect to the
${\mathbb{M}}^4$ metric. For each $p \in S$, the tangents at $p$
to those $\gamma^\pm_{\theta,\phi}\left(t \right)$ that pass
through $p$ can be shown to generate the null cone of $T_pS$.
Hence $S$ is a photon surface in the sense of Definition
\protect\ref{photonsurfacedef}. In terms of the double null
coordinates
\begin{align} u&:=t+r  \\ v&:= t-r
\end{align}
for $r:= (x^2+y^2+z^2)^{1/2}$, equation (\protect\ref{ellhyp4})
assumes the simple form
\begin{equation}
uv = -a^2 \ .
\end{equation}
In the future direction $S$ tends asymptotically to the null
hypersurface $\{ v=0\}$, whilst in the past direction $S$ tends
asymptotically to the null hypersurface $\{ u=0\}$. }
\end{example}
\begin{example}[De Sitter space] \label{deSitter}
{\rm De Sitter space-time may be regarded \protect\cite{HE73} as
a single-sheeted hyperboloid in Minkowski $5$-space
${\mathbb{M}}^5$.  By analogy with Examples 1 and 2, de Sitter
space-time is thus realized as a photon surface in
${\mathbb{M}}^5$.  }
\end{example}
\begin{example}[The Robertson-Walker models]\label{RobWal}
{\rm Since all Robertson-Walker models are conformally flat and
therefore locally conformally transformable to Minkowski space,
the photon surfaces of any such model may thus be obtained, at
least locally, by conformal transformations of Minkowski space. }
\end{example}

\begin{theorem}
\label{maintheorem} Let $S$ be a timelike hypersurface of
$\left(M,{\bf g}\right)$. Let  ${\bf n}$ be a unit normal field to
$S$ and let $h_{ab}$ be the induced metric on $S$. Let $\chi_{ab}$
be the second fundamental form on $S$ and let $\sigma_{ab}$ be the
trace-free part of $\chi_{ab}$. Then the following are equivalent:
\begin{enumerate}[i)]
\item $S$ is a photon  surface;
\item $\chi_{ab} k^a k^b =0$
$\forall$ null ${\bf k} \in T_pS$ $\forall p \in S$;
\item
$\sigma_{ab}=0$;
\item every affine null geodesic of
$(S,{\bf h})$ is an affine null geodesic of  $(M,{\bf g})$.
\end{enumerate}
\end{theorem}

\begin{proof} (i) $\Rightarrow$ (ii). Suppose $S$ is a photon
surface. Let $p \in S$ and let ${\bf k} \in T_p S$ be null. There
exists an affine null geodesic $ \gamma: \left(-\epsilon,\epsilon
\right) \rightarrow M $ of $\left(M,{\bf g}\right)$ such that
$\dot{\gamma}\left(0\right) = {\bf{k}}, |\gamma| \subset S$. One
has
\begin{equation}
   \chi_{ab} \dot{\gamma }^a \dot{\gamma}^b = n_{a;b} \dot{\gamma}^a
   \dot{\gamma}^b = \left(n_a \dot{\gamma}^a\right)_{;b} \dot{\gamma}^b = 0
   \label{eq8}
\end{equation}
along $\gamma$. At $p$ this gives $\chi_{ab} k^a k^b = 0$.

(ii) $\Rightarrow$ (iii).  Let $p \in S$.  By (ii) one has
$\sigma_{ab}k^ak^b=\chi_{ab}k^ak^b=0$ $\forall$ null ${\bf k}\in
T_pS$. Let $\{ {\bf e}_{(0)}, {\bf e}_{(1)}, {\bf e}_{(2)} \}$ be
an orthonormal basis for $T_pS$ with ${\bf e}_{(0)}$ timelike and
${\bf e}_{(1)}$, ${\bf e}_{(2)}$ spacelike. Any null ${\bf k}\in
T_pS$, normalized such that $g({\bf k},{\bf e}_{(0)})=-1$, has
components $k^a=(1,\cos \psi , \sin \psi )$ with respect to $\{
{\bf e}_{(0)}, {\bf e}_{(1)}, {\bf e}_{(2)} \}$ for some $\psi
\in [ 0,2\pi )$. A calculation gives
\begin{multline}
\sigma_{ab}k^ak^b = (\sigma_{00}+\smallfrac{1}{2}\sigma_{11}
+\smallfrac{1}{2}\sigma_{22})+ 2\sigma_{01}\cos \psi
+2\sigma_{02}\sin \psi
\\ + \smallfrac{1}{2}(\sigma_{11}-\sigma_{22})\cos 2\psi +
\sigma_{12}\sin 2\psi  \ .
\end{multline}
This must vanish for all $\psi \in [0,2\pi )$. One thus has
$\sigma_{01} =\sigma_{02}=\sigma_{12}=0$ and
$-\sigma_{00}=\sigma_{11}=\sigma_{22}$. Since $\sigma_{ab}$ is
trace-free one must also have
$\sigma_{00}=\sigma_{11}+\sigma_{22}$.  There follows
$\sigma_{ab}=0$.

(iii) $\Rightarrow$ (iv). For any curve in $S$ with null tangent
${\bf k}$ one has
\begin{equation}
k^a\dummy_{||b}k^b = h^a\dummy_c k^c\dummy_{;b}k^b =
k^a\dummy_{;b}k^b +(\sigma_{bc}k^bk^c)n^a \label{lgrad}
\end{equation}
where $||$ denotes covariant differentiation in $S$ with respect
to ${\bf h}$. The second term on the right of
$(\protect\ref{lgrad})$ vanishes by hypothesis. If ${\bf k}$ is
tangent to an affine null geodesic of $(S,{\bf h})$ then the term
on the left of $(\protect\ref{lgrad})$ also vanishes and so ${\bf
k}$ is tangent to an affine null geodesic of  $(M,{\bf g})$.

(iv) $\Rightarrow$ (i). Let $p \in S$ and let ${\bf k} \in T_p S$
be null. Let $\gamma : (-\epsilon,\epsilon ) \rightarrow S$ be an
affine null geodesic of $(S,{\bf h})$ such that $\dot{\gamma}(0) =
{\bf k}$. Then, by (iv), $\gamma$ is an affine null geodesic of
$(M,{\bf g})$ such that $\dot{\gamma}(0) = {\bf k}$, $|\gamma|
\subset S$.
\end{proof}

Condition (iii) of Theorem \protect\ref{maintheorem} is
equivalent to a requirement that $\chi_{ab}$ is pure trace in the
sense of
\begin{equation}
\chi_{ab} = \smallfrac{1}{3}\Theta  h_{ab} \ ,
\label{2ffsimplifies}
\end{equation}
where $\Theta := h^{cd}\chi_{cd}$ is the expansion of the unit
normal to $S$. For Example \protect\ref{Min3} one has $\Theta =
2/a$; for Example \protect\ref{Min4} one has $\Theta = 3/a$.
(Note that, by a standard abuse of notation, $h_{ab}$ denotes
both the induced metric on $S$ and the symmetric tensor field of
rank $(0,2)$ along $S$ in $M$ which satisfies $h_{ab}n^b=0$ and
pulls back to the induced metric on $S$.)

It is clear from condition (iii) of Theorem
\protect\ref{maintheorem} that a space-time  must be specialized
in some respect in order to admit any timelike photon surfaces in
the sense of Definition \protect\ref{photonsurfacedef}. For this
reason it is helpful to restrict attention to space-times which
admit groups of symmetries.

\begin{definition}\label{photonspheredef} {\rm Suppose $(M,{\bf g})$
admits a group $G$ of isometries.  A photon surface $S$ of
$(M,{\bf g})$ that is invariant under $G$, in the sense that each
$g \in G$ maps $S$ onto itself, will be called a $G$-invariant
photon surface.}
\end{definition}

Clearly any $G$-invariant null hypersurface is a $G$-invariant
photon surface.  In particular, if $G={\mathbb{R}}$ or
$G={\mathbb{S}}$, then any Killing horizon
\protect\cite{Car66,Boy69} is a $G$-invariant photon surface.
\protect\section{Dynamic Spherical symmetry: General
theory}\label{dssgt} By definition, a general spherically
symmetric space-time admits an $SO(3)$ isometry group for which
the group orbits are spacelike $2$-spheres. The following result
describes the evolution of the cross-sectional area of an
$SO(3)$-invariant photon surface in a spherically symmetric
space-time.
\begin{theorem} \label{firstphotonspheretheorem}
Let $(M,{\bf g})$ be a spherically symmetric space-time. Let $S$
be an $SO(3)$-invariant timelike hypersurface of $(M,{\bf g})$
and let ${\bf X}$ be the $SO(3)$-invariant unit future-directed
timelike tangent vector field along $S$ orthogonal to the
$SO(3)$-invariant $2$-spheres in $S$. Let $\mathscr{T}$ be one
such $SO(3)$-invariant $2$-sphere in $S$ and let $\mathscr{T}_s$
be the $SO(3)$-invariant $2$-sphere in $S$ at arc-length $s$ from
$\mathscr{T}$ along the integral curves of ${\bf X}$. Then $S$ is
a photon surface of $(M,{\bf g})$ iff the area $\pretwo \! A_s$
of $\mathscr{T}_s$ satisfies
\begin{equation}
\frac{d^2}{ds^2}\pretwo \! A_s = \frac{1}{4\pretwo \! A_s}\left(
\frac{d}{ds} \pretwo \! A_s \right)^2 +\pretwo \! A_s
\left(\smallfrac{1}{3}\Theta^2 -G_{ab}n^an^b \right) -4\pi
\label{2areaevolution}
\end{equation}
where $n^a$ is the unit normal to $S$, $\Theta$ is the expansion
of $n^a$ and $G_{ab}:= R_{ab}-\frac{1}{2}Rg_{ab}$ is the Einstein
tensor of $(M,{\bf g})$.
\end{theorem}

\begin{proof}  Let $h_{ab}$  be the induced Lorentzian $3$-metric
on $S$ and, for each $s$, let $\pretwo h_{ab}$ be the induced
Riemannian $2$-metric on $\mathscr{T}_s$.  The expansion of ${\bf
X}$ in $(S, {\bf h})$ is given by $\pretwo \Theta = \pretwo
h^{ab}X_{a;b}$ where the covariant derivative is that of  $(M,{\bf
g})$.  Since ${\bf X}$ is both shear-free and vorticity-free in
$(S,{\bf h})$, the Raychaudhuri equation for ${\bf X}$ in $(S,{\bf
h})$ assumes the form
\begin{equation}
\frac{d}{ds}\pretwo \Theta = -\smallfrac{1}{2}(\pretwo \Theta )^2
- \prethree \! R_{ab}X^aX^b \label{3dRay}
\end{equation}
where $\prethree \! R_{ab}$ is the Ricci tensor of $(S,{\bf h})$.

From first principles one has
\begin{align}
\pretwo \! R &= \prethree \! R + 2\prethree \! R_{ab}X^aX^b - (
\pretwo \chi^a\dummy_a )^2
+ \pretwo\chi^a\dummy_b \pretwo\chi^b\dummy_a \label{2R}\\
\prethree \! R &= R - 2R_{ab}n^an^b + \left(
\chi^a\dummy_a\right)^2 -\chi^a\dummy_b\chi^b\dummy_a \label{3R}
\end{align}
where $\pretwo\chi_{ab}$ is the second fundamental form of each
$\mathscr{T}_s$ in $(S,{\bf h})$.  Since $X^a$ is shear-free and
vorticity free in $(S,{\bf h})$ one has $\pretwo\chi_{ab}=
\frac{1}{2}\pretwo \Theta \pretwo h_{ab}$.  The second fundamental
form of $S$ admits the canonical decomposition $\chi_{ab}=
\frac{1}{3}\Theta h_{ab} +\sigma_{ab}$.  Equations
(\protect\ref{2R}) and (\protect\ref{3R}) therefore give
\begin{align}
\pretwo \! R &= \prethree \! R + 2\prethree \! R_{ab}X^aX^b -
\smallfrac{1}{2}
\pretwo\Theta^2 \\
\prethree \! R &= R -2R_{ab}n^an^b +\smallfrac{2}{3}\Theta^2
-\sigma^a\dummy_b\sigma^b\dummy_a
\end{align}
which combine to yield
\begin{equation}
2\prethree \! R_{ab}X^aX^b = \pretwo \! R + 2G_{ab}n^an^b
-\smallfrac{2}{3}\Theta^2 + \smallfrac{1}{2}(\pretwo \Theta )^2
+\sigma^a\dummy_b\sigma^b\dummy_a \ .
\end{equation}
One may now substitute for the second term on the right of
(\protect\ref{3dRay}) to obtain
\begin{equation}
\frac{d}{ds}\pretwo \Theta = -\smallfrac{3}{4}(\pretwo \Theta )^2
+\smallfrac{1}{3}\Theta^2 - \smallfrac{1}{2}\pretwo \! R -
G_{ab}n^an^b -\smallfrac{1}{2}\sigma^a\dummy_b\sigma^b\dummy_a \ .
\label{nearlythere}
\end{equation}
From first principles one has $\pretwo \Theta = \frac{d}{ds} \ln
\pretwo \! A$, and the Gauss-Bonnet theorem gives $\pretwo \!
R\pretwo \! A = 8\pi $. Substituting for $\pretwo \Theta$ and
$\pretwo \! R$ in (\protect\ref{nearlythere}) one obtains
\begin{equation}
\frac{d^2}{ds^2}\pretwo\! A_s = \frac{1}{4\pretwo\! A_s}\left(
\frac{d}{ds} \pretwo\! A_s \right)^2 + \pretwo\! A_s \left(
\smallfrac{1}{3}\Theta^2 -G_{ab}n^an^b -
\smallfrac{1}{2}\sigma^a\dummy_b\sigma^b\dummy_a \right) -4\pi \ .
\end{equation}
This agrees with (\protect\ref{2areaevolution}) iff
$\sigma^a\dummy_b\sigma^b\dummy_a =0$.

Construct, for the tangent bundle $TS$ of $S$, an orthonormal
basis field of the form $\{ {\bf X}, {\bf e}_{(1)}, {\bf e}_{(2)}
\}$, with ${\bf e}_{(1)}$ and ${\bf e}_{(2)}$ unit spacelike.
With respect to this basis one has
\begin{equation}
\sigma^a\dummy_b\sigma^b\dummy_a = (\sigma^0\dummy_0)^2
+(\sigma^1\dummy_1)^2 +(\sigma^2\dummy_2)^2
+2(\sigma^1\dummy_2)^2 -2(\sigma^1\dummy_0)^2
-2(\sigma^2\dummy_0)^2 \ .
\end{equation}
By spherical symmetry the vector field $\sigma^a\dummy_bX^b$ must
be proportional to $X^a$.  Hence one has
$\sigma^1\dummy_0=\sigma^2\dummy_0=0$. The vanishing of
$\sigma^a\dummy_b\sigma^b\dummy_a$ is thus equivalent to the
vanishing of $\sigma_{ab}$.  One has $\sigma_{ab}=0$ iff $S$ is a
photon surface. \end{proof}

A spherically symmetric metric is locally expressible in the form
\begin{equation}
 g_{ab} =  \left( \begin{array}{cccc}
   g_{00}   &  g_{01}       & 0                            & 0  \\
   g_{10}   &  g_{11}       & 0                            & 0  \\
   0        &  0            & g_{\theta \theta}            & 0  \\
   0        & 0             & 0               & g_{\theta \theta} \sin^2 \theta \\
\end{array}          \right)
\label{gss}
\end{equation}
with respect to coordinates $\left(x^0, x^1, \theta, \phi
\right)$ adapted to the spherical symmetry, where $g_{00}$,
$g_{01} = g_{10}$, $g_{11}$ and $g_{\theta \theta}>0$ depend only
on $x^0$ and $x^1$.  It is often convenient to introduce a radial
coordinate $r$, depending only on $x^0$ and $x^1$, such that
$g_{\theta \theta}$ is a function of $r$ only.  One is free to
specify $g_{\theta\theta}$ as a function of $r$ alone since to do
so is, in effect, a definition of the coordinate $r$. This will
be assumed to be done throughout this paper.

The following result is useful in the locating of
$SO(3)$-invariant photon surfaces in dynamic spherically
symmetric space-times.

\begin{lemma} \label{mylemma}
Let $(M,{\bf g})$ be a spherically symmetric space-time. Let $S$
be an $SO(3)$-invariant timelike hypersurface of $(M,{\bf g})$
and let ${\bf X}$ be the $SO(3)$-invariant unit future-directed
timelike tangent vector field along $S$ that is orthogonal to the
$SO(3)$-invariant $2$-spheres in $S$. Then $S$ is a photon
surface of $(M,{\bf g})$ iff
\begin{equation} X^a\dummy_{;b}X^b =
\smallfrac{1}{2}(g^{\theta\theta}n^b\partial_bg_{\theta\theta})n^a
\label{gradxx}
\end{equation}
holds along $S$, where $n^a$ is the unit normal field to $S$ in
$(M,{\bf g})$.
\end{lemma}

\begin{proof}  By spherical symmetry, and since ${\bf X}$ is unit
timelike, the vector field $\nabla_{\bf X}{\bf X}$ must be
proportional to ${\bf n}$. Hence it suffices to show that $S$ is
a photon surface iff along $S$ one has
\begin{equation} {\bf n \cdot}\nabla_{\bf X}{\bf X} =
\smallfrac{1}{2}g^{\theta\theta}n^b\partial_b g_{\theta\theta}
\end{equation}
or equivalently
\begin{equation}
\chi_{ab}X^aX^b = -\smallfrac{1}{2}g^{\theta\theta}n^a\partial_a
g_{\theta\theta} \ . \label{L1}
\end{equation}

Construct for $TS$ a local orthonormal basis field of the form
$\{ {\bf X}, {\bf e}_{(\theta )},{\bf e}_{(\phi )}\}$.  With
respect to this basis field the components of $\chi^a\dummy_b$
form a diagonal matrix with
\begin{equation}
\chi^{\theta}\dummy_{\theta} =
\chi^{\phi}\dummy_{\phi}=n^{\theta}\dummy_{;\theta} =
\smallfrac{1}{2}g^{\theta\theta}n^a\partial_a g_{\theta\theta} \ .
\end{equation}
Equation (\protect\ref{L1}) is thus equivalent to $\chi^0\dummy_0
= \chi^{\theta}\dummy_{\theta}=\chi^{\phi}\dummy_{\phi}$ which is
in turn equivalent to
$\sigma^0\dummy_0=\sigma^{\theta}\dummy_{\theta}=
\sigma^{\phi}\dummy_{\phi}$. In view of the trace-free property
of $\sigma^a \dummy_b$, equation (\protect\ref{L1}) is thus
equivalent to $\sigma^a \dummy_b=0$. From Theorem
\protect\ref{maintheorem} one has that $\sigma^a \dummy_b =0$
holds along $S$ iff $S$ is a photon surface. \end{proof}

Let us continue to work with respect to the coordinate system $\{
x^0,x^1,\theta , \phi \}$ employed in (\protect\ref{gss}). Let
$x^a (s)$ be an integral curve of the vector field ${\bf X}$ in
Lemma \protect\ref{mylemma}.  One has  \begin{equation}
\frac{dx^a}{ds} = X^a
\end{equation}
and equation $(\protect\ref{gradxx})$ becomes
\begin{equation}
\frac{d^2x^a}{ds^2}+\Gamma^a_{bc} \frac{dx^b}{ds} \frac{dx^c}{ds}
= \smallfrac{1}{2} (g^{\theta\theta}n^b\partial_b
g_{\theta\theta})n^a \ . \label{x2s}
\end{equation}
Since ${\bf X} = \frac{dx^0}{ds} \left(1, \frac{dx^1}{dx^0}, 0, 0
\right)$ is unit timelike one has
\begin{equation}
\left(\frac{ds}{dx^0}\right)^2 = - g_{ab} \frac{dx^a}{dx^0}
\frac{dx^b}{dx^0} \ .
\end{equation}
The $x^0$ and $x^1$ components of equation $(\protect\ref{x2s})$
thus combine to give
\begin{equation}
\frac{d^2x^1}{(dx^0)^2} = - \
\smallfrac{1}{2}g^{\theta\theta}n^a\partial_ag_{\theta\theta}
\left(n^1   - \frac{dx^1}{dx^0} n^0 \right) g_{bc}
\frac{dx^b}{dx^0} \frac{dx^c}{dx^0}  + \left(\frac{dx^1}{dx^0}
\Gamma^0_{ab} - \Gamma^1_{ab} \right) \frac{dx^a}{dx^0}
\frac{dx^b}{dx^0}  \label{x12x0}
\end{equation}
where the components of ${\bf n}$ are given by
\begin{equation}
n^0 = \psi \  g_{1a}\  \frac{dx^a}{dx^0}; \quad n^1 = - \psi \
g_{0a}\   \frac{dx^a}{dx^0}
\end{equation}
for
\begin{equation}
\psi := \left(- \Delta \right)^{- \frac{1}{2}}
    \left(-g_{ab} \frac{dx^a}{dx^0}\frac{dx^b}{dx^0}\right)^{- \frac{1}{2}}
\end{equation}
where
\begin{equation}\label{Deltadef}
\Delta:= g_{00}g_{11}-(g_{01})^2
\end{equation}
is the determinant of the time-space part of $g_{ab}$ in
(\protect\ref{gss}). Equation (\protect\ref{x12x0}) is the
coordinate equivalent of (\protect\ref{gradxx}) and provides for
the easy determination of $SO(3)$-invariant photon surfaces (see
Example \protect\ref{Vaidyaex}).
\protect\section{Static Spherical Symmetry: General Theory} By
definition, a spherically symmetric space-time is static if it
admits an $SO(3)\times {\mathbb{R}}$ group of isometries such that the
${\mathbb{R}}$ orbits are generated by a Killing field ${\bf K}$
which is both hypersurface orthogonal and orthogonal to the
$SO(3)$ orbits.  The present section will be concerned with
$SO(3)\times {\mathbb{R}}$-invariant photon surfaces in static
spherically symmetric space-times.  Such surfaces may be termed
photon spheres because, as will be seen, they are a natural
generalization to general static spherically symmetric space-times
of the Schwarzschild photon sphere concept. The term ``photon
sphere'' will be regarded as applicable only in static
spherically symmetric space-times. For clarity, the term
``$SO(3)\times {\mathbb{R}}$-invariant photon surface'' will usually be
employed in preference to ``photon sphere''.

Although the space-times of Examples \protect\ref{Min3} and
\protect\ref{Min4} are static and spherically symmetric, the
photon surfaces in these space-times are not
$SO(3)\times {\mathbb{R}}$-invariant and so are not photon spheres.  Of
the Robertson-Walker space-times of Example \protect\ref{RobWal},
only the Einstein cylinder is both spherically symmetric and
static.  None of the photon surfaces of the Einstein cylinder are
$SO(3)\times {\mathbb{R}}$-invariant.  Thus the Einstein cylinder has no
photon spheres.

One may characterize an $SO(3)\times {\mathbb{R}}$-invariant photon
surface, or photon sphere, in a static spherically symmetric
space-time by means of the following special case of Theorem
\protect\ref{firstphotonspheretheorem}.

\begin{theorem}\label{statssgeomtheorem}
Let $(M,{\bf g})$ be a static spherically symmetric space-time
with Killing field ${\bf K}$ and let $S$ be an
$SO(3)\times {\mathbb{R}}$-invariant timelike hypersurface of $(M,{\bf
g})$.  Then $S$ is an $SO(3)\times {\mathbb{R}}$-invariant photon
surface of $(M,{\bf g})$ if there exists an $SO(3)$-invariant
$2$-sphere $\mathscr{T} \subset S$ satisfying
\begin{equation} A\left( \Theta^2 - 3G_{ab}n^an^b \right) = 12\pi
\label{mainequation}
\end{equation}
where $A$ is the area of $\mathscr{T}$, $n^a$ is the unit normal
to $S$ and $\Theta$ is the trace of the second fundamental form of
$S$.  Conversely, if $S$ is an $SO(3)\times {\mathbb{R}}$-invariant
photon surface of $(M,{\bf g})$ then (\protect\ref{mainequation})
holds for every $SO(3)$-invariant $2$-sphere $\mathscr{T}\subset
S$.
\end{theorem}

\begin{proof}  Note that the unit future-directed timelike tangent
field ${\bf X}$ along $S$ in Theorem
\protect\ref{firstphotonspheretheorem} is proportional to the
restriction to $S$ of the Killing field ${\bf K}$.

Suppose first that there exists an $SO(3)$-invariant $2$-sphere
$\mathscr{T}\subset S$ such that (\protect\ref{mainequation})
holds. The quantities $A$, $\Theta$ and $G_{ab}n^an^b$ remain
constant as $\mathscr{T}$ is mapped along the flow lines of the
Killing field ${\bf K}$. So they also remain constant as they are
mapped along the flow lines of ${\bf X}$.  Hence
(\protect\ref{2areaevolution}) holds with the term on the left
and the first term on the right both zero. Thus, by Theorem
\protect\ref{firstphotonspheretheorem}, $S$ is a photon surface of
$(M,{\bf g})$. By hypothesis $S$ is
$SO(3)\times {\mathbb{R}}$-invariant.

For the converse, suppose that $S$ is an
$SO(3)\times {\mathbb{R}}$-invariant photon surface of $(M,{\bf g})$.
Then (\protect\ref{2areaevolution}) holds for every
$SO(3)$-invariant $2$-sphere $\mathscr{T}_s\subset S$.  Since
${\bf K}$ induces groups of local isometries, the area $A_s$ of
$\mathscr{T}_s$ is independent of the parameter $s$. Hence the
term on the left and the first term on the right of
(\protect\ref{2areaevolution}) both vanish and one obtains
(\protect\ref{mainequation}). \vspace{-3ex}\end{proof}

\begin{corollary*} If one has $G_{ab}Y^aY^b\geq 0$ $\forall$ vectors ${\bf Y}$
and $S$ is an $SO(3)\times {\mathbb{R}}$-invariant timelike photon
surface of $(M,{\bf g})$, then for any $SO(3)$-invariant
$2$-sphere $\mathscr{T}\subset S$ one has \begin{equation}
A\Theta^2 \geq 12\pi \label{mainequationcor}\end{equation} with
equality holding iff $G_{ab}n^an^b=0$ along $S$. \noproof
\end{corollary*}

For Schwarzschild space-time where (see Example
\protect\ref{Schwarzschildex}) the only timelike photon sphere is at
$r=3m$,  one has $A=4\pi (3m)^2$, $\Theta = 1/(\sqrt{3}m)$ and
$G_{ab}=0$ which verifies (\protect\ref{mainequation}) and
(\protect\ref{mainequationcor}) for this case.

If the Einstein equations hold with a zero cosmological constant
then, in the corollary to Theorem
\protect\ref{statssgeomtheorem}, the hypothesis $G_{ab}Y^aY^b
\geq 0$ for all vectors ${\bf Y}$  is equivalent to
$T_{ab}Y^aY^b\geq 0$ for all vectors ${\bf Y}$. This is a
physically reasonable energy condition.  In particular, for a
perfect fluid with density $\rho$ and pressure $p$, it is
equivalent to a condition that $\rho$ and $p$ are both
non-negative.  More generally, the condition holds for an energy
tensor with a single timelike eigenvector (type I in the
classification of Hawking \& Ellis \protect\cite{HE73}) iff each
energy tensor eigenvalue is non-negative.

The characterization of timelike photon surfaces provided by Theorem
\protect\ref{statssgeomtheorem} involves derivatives of the
metric components up to second order. The following result
(Theorem \protect\ref{genphotonspheretheorem}) provides an
entirely different characterization of
$SO(3)\times{\mathbb{R}}$-invariant photon surfaces in terms of
derivatives of the metric components up to only first order.

Let a general static spherically symmetric metric ${\bf g}$ be
expressed in the form (\protect\ref{gss}) with $g_{\theta\theta}$
a function of $r$ only.  The Killing equation $K_{(a;b)}=0$ and
the orthogonality of ${\bf K}$ to $\partial_{\theta}$ and
$\partial_{\phi}$ gives $K^a\partial_a g_{\theta\theta} =0$ and
hence $\nabla_{\bf K}r=0$ where $r$ is to be regarded as a scalar
field on $M$. Since $r$ is independent of $\theta$ and $\phi$ it
follows that any $SO(3)\times {\mathbb{R}}$-invariant hypersurface $S$
of $(M,{\bf g})$ must be of the form $\{r=\text{\em const.}\}$. If
$S$ is also a timelike hypersurface then $r^{;a}$ is a spacelike
vector field along $S$ and ${\bf K}$ is a timelike vector field in
a neighbourhood of $S$.

If the coordinates $x^0$ and $x^1$ implicit in
(\protect\ref{gss}) are chosen such that ${\bf K}=\partial_{x^0}$
then the Killing equation gives that all the metric components
$g_{ab}$ in (\protect\ref{gss}) are independent of $x^0$. Since
$r$ is constant along the integral curves of ${\bf K}$, the
coordinate $x^1$ must be a function of $r$ only.  A natural
choice is $x^1=r$.  One may redefine $x^0$ according to
$x^0\rightarrow x^0-\int(g_{01}/g_{00})dx^1$. This diagonalizes
the time-space part of $g_{ab}$ and leaves the components of
$g_{ab}$ independent of $x^0$. Furthermore the curves $\{
x^1,\theta , \phi =\text{\em const.}\}$ are unchanged except that
they are re-parametrized. The vector field $\partial_{x^0}$ then
becomes a conformal Killing field.

Define the tensor field
\begin{equation}
\epsilon^{ab} := (-\Delta )^{-1/2} \left( \begin{array}{cccc}
0&1&0&0\\ -1&0&0&0 \\ 0&0&0&0 \\ 0&0&0&0 \end{array} \right)
\end{equation}
on $M$, where the components are given with respect to the
coordinate basis employed in (\protect\ref{gss}) and $\Delta$ is
the determinant of the time-space part of ${\bf g}$ in
(\protect\ref{gss}), as in (\protect\ref{Deltadef}).
\begin{theorem} \label{genphotonspheretheorem} Let $(M,{\bf g})$ be a static
spherically symmetric space-time with ${\bf g}$ of the form
(\protect\ref{gss}), with $g_{\theta\theta}$ a function of the
coordinate $r$ only. Let $S$ be an $SO(3)\times {\mathbb{R}}$-invariant
timelike hypersurface of $(M,{\bf g})$ and suppose that $\nabla
r$ is nowhere-zero along $S$. Then $S$ is an
$SO(3)\times {\mathbb{R}}$-invariant photon surface of $(M,{\bf g})$ iff
\begin{equation}
2g_{\theta\theta}\epsilon^{ab}\epsilon^{cd}r_{;ac}r_{;b}r_{;d}
+r^{;a}r_{;a}r^{;c}\partial_{c}g_{\theta\theta}=0
\label{genphotonsphere} \end{equation} holds along $S$.
\end{theorem}

\begin{proof}  Since $(M,{\bf g})$ is both spherically symmetric
and static, the surface $S$ is of the form $\{ r=\text{\em
const.}\}$. The unit spacelike normal to $S$ is therefore given
by  \begin{equation} n^a = \eta r^{;a}
\end{equation}
for
\begin{equation}
\eta := (r^{;a}r_{;a})^{-1/2}\ .
\end{equation}
The second fundamental form of $S$ is given by
\begin{equation}
\chi_{ab}:= \eta h_a\Dummy^ch_b\Dummy^d r_{;cd} \ .
\end{equation}
The vector fields
\begin{align}
{\bf X}^a &:= (-\Delta g^{bc}r_{;b}r_{;c})^{-1/2}(r_{;1},-r_{;0},0,0)\\
{\bf e}_{(\theta )} &:= (g_{\theta\theta})^{-1/2}\partial_{\theta}\\
{\bf e}_{(\phi )} &:= (g_{\theta\theta}\sin^2\theta
)^{-1/2}\partial_{\phi}
\end{align}
form an orthonormal frame field along $S$, with ${\bf e}_{(\theta
)}$ and ${\bf e}_{(\phi )}$ unit spacelike and ${\bf X}$ unit
timelike. One has
\begin{equation}
\chi_{ab}X^a{\bf e}_{(\theta )}^b = \chi_{ab}X^a{\bf e}_{(\phi
)}^b =\chi_{ab}{\bf e}_{(\theta )}^a{\bf e}_{(\phi )}^b=0
\end{equation}
\begin{equation}
\chi_{ab}{\bf e}_{(\theta )}^a{\bf e}_{(\theta )}^b=
\chi_{ab}{\bf e}_{(\phi )}^a {\bf e}_{(\phi )}^b = \frac{\eta
r^{;a}\partial_{a}g_{\theta\theta}}{2g_{\theta\theta}}
\end{equation}
\begin{equation}
\chi_{ab}X^a X^b = \eta^3
\epsilon^{ab}\epsilon^{cd}r_{;ac}r_{;b}r_{;d} \ .
\end{equation}
Condition (iii) of Theorem \protect\ref{maintheorem} holds iff
$\chi_{ab}$ is proportional to $h_{ab}$ and hence iff
\begin{equation}-\chi_{ab}X^a X^b =
\chi_{ab}{\bf e}_{(\theta )}^a{\bf e}_{(\theta )}^b =
\chi_{ab}{\bf e}_{(\phi )}^a{\bf e}_{(\phi )}^b \ .
\end{equation}
This is equivalent to
\begin{equation}-\eta^2 \epsilon^{ab}\epsilon^{cd}r_{;ac}r_{;b}r_{;d}
=\frac{r^{;a}\partial_a g_{\theta\theta}}{2g_{\theta\theta}}
\end{equation}
which is in turn equivalent to
(\protect\ref{genphotonsphere}).\end{proof}

Equation (\protect\ref{genphotonsphere}) can have solutions such
that $\{ r=\text{\em const.}\}$ is a spacelike hypersurface and
therefore not a photon surface (see e.g.\ Example
\protect\ref{rnex}). It is therefore always necessary in the use
of Theorem \protect\ref{genphotonspheretheorem} to check that the
hypersurface $\{ r=\text{\em const.}\}$ is in fact timelike or
null.

Note that in a region of space-time where the Killing field ${\bf
K}$ is spacelike, for example behind the event horizon of
Schwarzschild space-time, the hypersurfaces $\{ r=\text{\em
const.}\}$ are necessarily spacelike and so cannot be photon
spheres.\vspace{1ex}

{\bf Case 1.  ($x^1:= r$, components of $g_{ab}$ independent of
$x^0$.)} As discussed previously, for a static spherically
symmetric metric it is possible to choose $x^1:=r$, with
$g_{\theta\theta}$ depending only upon $r$ and with all the
components of $g_{ab}$ independent of $x^0$. In this case
(\protect\ref{genphotonsphere}) reduces to
\begin{equation}
g_{00}\partial_r g_{\theta\theta} = g_{\theta\theta}\partial_r
g_{00} \ . \label{photonsphere}
\end{equation}
This agrees with an equation obtained by Virbhadra \& Ellis
\protect\cite{VE99b} on the basis of a different definition
\protect\cite{VE99a} of a photon sphere. Note that even though
the components $g_{rr}, g_{0r}=g_{r0}$ do not appear in
(\protect\ref{photonsphere}), they are not assumed to vanish. A
particular sub-case of interest is that of time-space
coordinates, with $t:=x^0$ timelike in the sense of $g^{tt}< 0$.
Another sub-case of interest is that of single null (radiation)
coordinates, with $u:=x^0$ null in the sense of $g^{uu}=0$.\vspace{1ex}

{\bf Case 2. (Double null coordinates $u, v$.)} Let $x^0 := u$,
$x^1 :=v$ be double null coordinates in the sense of
$g^{uu}=g^{vv}=0$.  The radial coordinate $r$ is to be regarded
as a function of $u$ and $v$.  Then
(\protect\ref{genphotonsphere}) assumes the form
\begin{equation}
g_{\theta\theta}\{ r_{;uu} (r_{;v})^2-2r_{;uv}r_{;u}r_{;v}
+r_{;vv}(r_{;u})^2\}
+2(r_{;u}r_{;v})^2\partial_rg_{\theta\theta}=0 \ .
\label{photonspheredoublenull}
\end{equation}
The metric components $g_{uv}=g_{vu}$ enter here through the
covariant derivatives of $r$.\vspace{1ex}

Equations (\protect\ref{genphotonsphere}),
(\protect\ref{photonsphere}) and
(\protect\ref{photonspheredoublenull}) may be referred to as
photon sphere equations since they give the location of timelike photon
spheres in static spherically symmetric space-times.

In order to facilitate further progress, a general static
spherically symmetric metric will be written in such a form as to
cast the Einstein tensor in a particularly simple and convenient
form.

One has from previous remarks that a general static spherically
symmetric metric is locally expressible in the form
\begin{equation}
{\bf g} = g_{tt}dt^2 +g_{rr}dr^2 + r^2(d\theta^2 +\sin^2\theta
d\phi^2 )
\end{equation}
where $g_{tt}$ and $g_{rr}$ are functions of $r$ only.  Let
\begin{align}
m(r) &:= \smallfrac{1}{2}r \left( 1 - \frac{1}{g_{rr}}\right) \\
\mu (r)&:= \ln (-g_{tt}g_{rr}) \ .
\end{align}
Then the metric assumes the form
\begin{equation}
{\bf g} = -\left( 1-\frac{2m(r)}{r}\right){\rm e}^{\mu (r)} dt^2
+\left(1-\frac{2m(r)}{r}\right)^{-1}dr^2 +r^2(d\theta^2 +
\sin^2\theta d\phi^2 ) \label{metricform}
\end{equation}
and the Einstein tensor is given by
\begin{equation}
G^a\dummy_b = 8\pi\left( \begin{array}{cccc} -\rho (r) & 0 &0&0\\
0&p_1(r)&0&0 \\ 0&0& p_2(r) &0\\ 0&0&0&p_2(r)
\end{array} \right)
\end{equation}
for
\begin{align}
8\pi \rho (r) &:= \frac{2m'(r)}{r^2} \label{den}\\
8\pi p_1(r) &:= \frac{1}{r^2}\left\{ (r-2m(r))\mu
'(r)-2m'(r)\right\}
\label{pr}\\
8\pi p_2(r)&:= \frac{1}{4r^2}\left\{ (2(r+m(r)-3rm'(r))\mu '(r)
+r(r-2m(r))(\mu '(r))^2 \right. \nonumber \\
& \hspace*{12em}\left. -4rm''(r) +2r(r-2m(r))\mu '' (r) \right\}
\ . \label{ptheta} \end{align} where a prime denotes
differentiation with respect to $r$. For the right sides of
(\protect\ref{den}), (\protect\ref{pr}) and (\protect\ref{ptheta})
to be defined at some radius $\hat{r}>0$ one evidently needs
$m(r)$ and $\mu (r)$ to be twice differentiable at $r=\hat{r}$.
Equations (\protect\ref{den}) and (\protect\ref{pr}) combine to
give
\begin{equation}\mu '(r) =
\frac{8\pi r(\rho (r) +p_1(r))}{\left( 1 - \frac{2m(r)}{r}
\right)} \label{lambdadash}
\end{equation}
whereby one may rewrite (\protect\ref{ptheta}) in the more
convenient form
\begin{equation}\frac{2}{r}(p_2(r)-p_1(r))=
p_1'(r)+\frac{ \left( \frac{m(r)}{r^2}+4\pi r
p_1(r)\right)\left(\rho (r)+p_1(r)\right)} {\left(
1-\frac{2m(r)}{r}\right) } \ .\label{pressurediff}
\end{equation}
In the perfect fluid case $p_1(r)=p_2(r)=:p(r)$ equation
(\protect\ref{pressurediff}) reduces to the
Tolman-Oppenheimer-Volkoff equation
\begin{equation}\label{TOV}
p'(r)=-\frac{ \left( \frac{m(r)}{r^2}+4\pi r
p(r)\right)\left(\rho (r)+p(r)\right)} {\left(
1-\frac{2m(r)}{r}\right) } \ .
\end{equation}
By means of equations (\protect\ref{den}) and
(\protect\ref{lambdadash}), the photon sphere equation
(\protect\ref{photonsphere}) becomes
\begin{equation}
1-\frac{3m(r)}{r}-4\pi r^2p_1(r)=0 \ . \label{fluidphotonsphere}
\end{equation}
For $r$ such that $2m(r)<r$, and hence such that the hypersurface
$\{ r=\text{\em const.}\}$ is timelike, equation
(\protect\ref{fluidphotonsphere}) gives the location of the
$SO(3)\times {\mathbb{R}}$-invariant timelike photon surfaces for the
metric (\protect\ref{metricform}).

Equation (\protect\ref{fluidphotonsphere}) is the basis for the
following result which shows that, subject to a suitable energy
condition, any black hole in a static spherically symmetric
space-time must be surrounded by an $SO(3)\times {\mathbb{R}}$-invariant
photon surface. For the purpose of this and subsequent results, a
function $f: {\mathbb R}\supset I \rightarrow {\mathbb R}$ on an
interval $I$ will be said to be piecewise $C^r$ if $I$ is the
disjoint union of a locally finite collection of intervals $I_i$
such that $f|I_i$ is $C^r$.  Each interval $I_i$ may be open,
closed or half-open.
\begin{theorem}\label{trappedsurfacetheorem}
Suppose the metric ${\bf g}$ has the form
(\protect\ref{metricform}) for $r_0<r<\infty$, for some $r_0>0$,
with $m(r)$ and $\mu (r)$ both $C^0$, piecewise $C^2$ functions
of $r \in (r_0, \infty )$. Suppose the following hold:
\begin{enumerate}[1)]
\item $\rho (r)$ and $p_1(r)$ are bounded functions of $r \in (r_0,\infty)$;
\item $2m(r)<r$ $\forall r \in (r_0,\infty)$;
\item $\rho (r) \geq 0$, $p_1(r)\geq 0$ $\forall r \in (r_0,\infty)$;
\item $\lim_{r \rightarrow \infty}4\pi r^2p_1(r)
=\lim_{r \rightarrow \infty}4\pi r^2\rho (r)=0$;
\item for each value of $t$ the $2$-surfaces $\mathscr{T}_{t,r}:=
\{ t=\text{\em const.}\} \cap \{ r=\text{\em const.} \}$,
$r_0<r<\infty$, are such that $\mathscr{T}_t := \lim_{r \searrow
r_0} \mathscr{T}_{t,r}$ exists as an embedded spacelike
$2$-sphere in $(M,{\bf g})$ and is marginally outer trapped.
\end{enumerate} Then $(M,{\bf g})$ admits an
$SO(3)\times {\mathbb{R}}$-invariant timelike photon surface of the form
$\{ r=r_1\}$ for some $r_1\in (r_0,\infty)$.
\end{theorem}
\begin{proof}  Fix $t$ and let ${\bf k}$ be the outward
future-directed null normal field along each $\mathscr{T}_{t,r}$,
$r_0<r<\infty$, normalized such that ${\bf g}({\bf k},{\bf n})=1$,
where ${\bf n}=\left( 1 - \frac{2m(r)}{r}\right)^{1/2}\partial_r$
is the outward radial unit tangent to $\{ t= \text{\em const.}\}$.
Since ${\bf k}$ is parallelly propagated along each of the
geodesic integral curves of ${\bf n}$, one has that $\lim_{r
\searrow r_0}{\bf k}$ is a well-defined, nowhere-zero null vector
field along $\mathscr{T}_t:= \lim_{r \searrow r_0}{\mathcal
T}_{t,r}$. For $r \in (r_0,\infty )$ the vector field ${\bf k}$
has the form $k^a = (k^t,a(r)k^t,0,0)$ for
\begin{equation}
a(r):=(-g_{tt}/g_{rr})^{1/2}=\left( 1-\frac{2m(r)}{r}\right){\rm
e}^{\mu (r)/2} \ .
\end{equation}
The expansion of ${\bf k}$ is given by
\begin{equation}
\Theta_{\text{\em out}} = \pretwo h^b \dummy_a k^a \dummy_{;b} =
\frac{2a(r)}{r}k^t \ . \end{equation} The condition that
$\mathscr{T}_t$ is marginally outer trapped therefore implies
\begin{equation}\label{limit}
0=\lim_{r \searrow r_0}a(r) = \lim_{r \searrow r_0}\left( 1 -
\frac{2m(r)}{r}\right) {\rm e}^{\mu (r)/2} \ .
 \end{equation}
The non-negativity of $\rho (r)$ and $p_1(r)$ gives, by means of
(\protect\ref{den}) and (\protect\ref{lambdadash}), that $m(r)$
and $\mu (r)$ are non-decreasing functions of $r \in
(r_0,\infty)$. Thus (\protect\ref{limit}) holds iff at least one
of
\begin{equation}\lim_{r \searrow r_0}\left( 1 - \frac{2m(r)}{r} \right) = 0 \ ;
\quad \lim_{r \searrow r_0}\mu (r) = -\infty \label{limitpair}
\end{equation}
holds.

Suppose the first of (\protect\ref{limitpair}) fails. Then the
second must hold and one has
\begin{equation}\lim_{r \searrow r_0}\left( 1 -\frac{2m(r)}{r}\right)^{-1} <\infty \ .
\label{finite}
\end{equation}
From the boundedness of $\rho (r)$ and $p_1(r)$ on $r \in
(0,\infty )$ one has, by means of (\protect\ref{lambdadash}) and
(\protect\ref{finite}), that $\limsup_{r \searrow r_0}\mu '(r)$
is finite. This is incompatible with the second of
(\protect\ref{limitpair}). Hence the first of
(\protect\ref{limitpair}) must hold.

Let $f: (r_0,\infty )\rightarrow {\mathbb R}$ be the left side of
(\protect\ref{fluidphotonsphere}). By the non-negativity of
$p_1(r)$ and the first of (\protect\ref{limitpair}) one has
$\lim_{r \searrow r_0}f(r) \leq -\frac{1}{2}$.  By condition (4),
equation (\protect\ref{den}) and l'H{\^o}pital's rule one has
$\lim_{r \rightarrow \infty}m(r)/r = \lim_{r \rightarrow \infty}
r^2p_1(r)=0$ and hence $\lim_{r \rightarrow \infty}f(r)=1$.
Hence there exists some $r_1\in (r_0,\infty)$ such that
$f(r_1)=0$.  The hypersurface $\{ r=r_1\}$ is an
$SO(3)\times {\mathbb{R}}$-invariant photon surface of $(M,{\bf g})$.
\end{proof}

Condition (3) of Theorem \protect\ref{trappedsurfacetheorem} may
be expressed more succinctly as $G_{ab}Y^aY^b\geq 0$ $\forall$
vectors ${\bf Y}$. With regard to condition (5) of Theorem
\protect\ref{trappedsurfacetheorem}, to have required ${\mathcal
T}_t$ to be contained in the hypersurface $\{ t=\text{\em
const.}\}$ would have been too strong since, for Schwarzschild
space-time, no spacelike hypersurface of the form $\{ t=\text{\em
const.}\}$ in the exterior region contains a marginally outer
trapped $2$-surface.

Theorem \protect\ref{trappedsurfacetheorem} may be interpreted to
the effect that, subject to the energy conditions expressed in
condition (3), any static spherically symmetric black hole must
be surrounded by an $SO(3)\times {\mathbb{R}}$-invariant timelike photon
surface. The following result may then be regarded as a partial
converse in that it shows, subject to a suitable energy
condition, that if there exists an
$SO(3)\times{\mathbb{R}}$-invariant timelike photon surface then there must
be a naked singularity or a black hole, or more than a certain
amount of matter.

\begin{proposition}\label{nopsprop}
Suppose the metric ${\bf g}$ has the form
(\protect\ref{metricform}) for $0 < r< \infty$, with $m(r)$ and
$\mu (r)$ both $C^0$, piecewise $C^2$ functions of $r\in
(0,\infty )$. If the following all hold:
\begin{enumerate}[1)]
\item $\rho (r)$ is a non-increasing, bounded function of $r \in (0,\infty)$;
\item $\lim_{r \rightarrow 0}m(r)=0$;
\item $4m(r)<r$ $\forall r \in (0,\infty)$;
\item $p_1(r) \leq \rho (r)/3$ $\forall r \in (0,\infty)$,
\end{enumerate}
then $(M,{\bf g})$ can contain no $SO(3)\times {\mathbb{R}}$-invariant
timelike photon surfaces.
\end{proposition}
\begin{proof} By conditions (1) and (2) with equation
(\protect\ref{den}) one has $m(r)\geq (4\pi /3)r^3\rho (r)$
$\forall r > 0$. By condition (4) one therefore has $4\pi
r^2p_1(r)\leq (4\pi /3)r^2\rho (r) \leq m(r)/r$ $\forall r>0$.
The left side of (\protect\ref{fluidphotonsphere}) is thus
bounded from below by $1-4m(r)/r$ $\forall r>0$.  This is
positive by condition (3). The left side of
(\protect\ref{fluidphotonsphere}) is therefore non-vanishing for
all $r>0$. \end{proof}

Note that this result is valid even for negative $p_1(r)$ and
$\rho (r)$. Condition (2) of Proposition \protect\ref{nopsprop}
prohibits any curvature singularity at $r=0$. Condition (3) may
be interpreted as a requirement that there is no black hole and
less than a certain amount of matter. The result shows that one
of these two conditions must fail if there is an
$SO(3)\times{\mathbb{R}}$-invariant timelike photon surface and conditions
(1) and (4) both hold.

When the matter is a perfect fluid it is possible to improve
condition (3) of Proposition \protect\ref{nopsprop} to condition
(3) of the following result.

\begin{theorem}\label{nopstheorem}
Suppose the metric ${\bf g}$ has the form
(\protect\ref{metricform}) for $0 < r< \infty$, with $m(r)$ and
$\mu (r)$ both $C^1$, piecewise $C^2$ functions of $r\in
(0,\infty )$. If the following all hold:
\begin{enumerate}[1)]
\item the matter is a perfect fluid with pressure $p(r)$ and
density $\rho (r)$;
\item $\lim_{r\rightarrow\infty}4\pi r^2p(r)
=\lim_{r\rightarrow\infty}4\pi r^2\rho (r) =0$;
\item $(24/7)m(r)<r$ $\forall r \in (0,\infty)$;
\item $p(r) \leq \rho (r)/3$ $\forall r \in (0,\infty)$,
\end{enumerate}
then $(M,{\bf g})$ can contain no $SO(3)\times {\mathbb{R}}$-invariant
timelike photon surfaces.
\end{theorem}
\begin{proof} Let $f:(0,\infty )\rightarrow {\mathbb{R}}$ be the
left side of equation (\protect\ref{fluidphotonsphere}).  Since
$m(r)$ and $\mu (r)$ are $C^1$, piecewise $C^2$ functions of $r\in
(0,\infty )$, one has by equation (\protect\ref{pr}) that $f(r)$
is a $C^0$, piecewise $C^1$ function of $r\in (0,\infty )$. The
function $f'(r)$ is then a piecewise $C^0$ function of $r\in
(0,\infty )$ which, by means of of the Tolman-Oppenheimer-Volkoff
equation (\protect\ref{TOV}), is given by
\begin{equation}\label{feqn}
rf'(r) = 3\frac{m(r)}{r}-12\pi r^2\rho (r) -8\pi r^2 p(r)
+\frac{\left(\frac{m(r)}{r}+4\pi r^2p(r)\right)}{\left(
1-2\frac{m(r)}{r}\right)}4\pi r^2 (\rho (r) +p(r)) \ .
\end{equation}
For $r=r_1\in (0,\infty )$ such that
\begin{equation}\label{fiszero}
0=f(r_1)=1-3\frac{m(r_1)}{r_1}-4\pi r_1^2p(r_1)
\end{equation}
equation (\protect\ref{feqn}) reduces to
\begin{equation}\label{feqn1}
r_1 f'(r_1)= 1-8\pi r_1^2 (\rho (r_1) +p(r_1) ) \ .
\end{equation}
From condition (3) and equation (\protect\ref{fiszero}) one has
$4\pi r_1^2 p(r_1)>1/8$, whence by condition (4) one has $4\pi
r_1^2\rho (r_1)>3/8$. Thus (\protect\ref{feqn1}) gives
$f'(r_1)<0$.

By condition (2), equation (\protect\ref{den}) and l'H{\^o}pital's
rule one has $\lim_{r\rightarrow\infty}m(r)/r =
\lim_{r\rightarrow\infty} 4\pi r^2p(r)=0$ and hence
$\lim_{r\rightarrow\infty}f(r)=1$. Since it has been established
that $f'(r)$ is negative for all $r\in (0,\infty )$ such that
$f(r)=0$, one must therefore have $f(r)>0$ for all $r\in
(0,\infty )$.  Hence the space-time can contain no
$SO(3)\times {\mathbb{R}}$-invariant timelike photon surfaces. \end{proof}

Note that, as for Proposition \protect\ref{nopsprop}, Theorem
\protect\ref{nopstheorem} is valid even for negative pressure and
density.

One would like to remove the insufficient matter parts of
condition (3) of Proposition \protect\ref{nopsprop} and condition
(3) of Theorem \protect\ref{nopstheorem}, in other words to
weaken these to a no-black-hole condition $2m(r)<r$ $\forall
r>0$. But no result to this effect is forthcoming. On the other
hand no counterexample is known.

To conclude this section it will be shown that the physical
significance of the photon sphere in Schwarzschild space-time, as
discussed in the Introduction, carries over to the general static
spherically symmetric case.  Suppose the metric has the form
(\protect\ref{metricform}) for $r_0 <r<\infty$ and is
asymptotically flat in the limit $r \rightarrow \infty$. Assume
$p_1(r)\geq 0$, $m(r)\geq 0$ $\forall r > r_0$. The matter need
not be a perfect fluid. Denote the left side of
(\protect\ref{fluidphotonsphere}) by $f(r)$. The condition of
asymptotic flatness gives $\lim_{r\rightarrow \infty}f(r) =1$ so,
if there are any $SO(3)\times{\mathbb{R}}$-invariant timelike
photon surfaces, there will be an outermost such surface $S$. For
simplicity assume $f'(r)\neq 0$ at $S$. Let ${\mathcal
R}_{\text{\em ext}}$ be the connected component of $\{ q \in M:
f(q)>0\}$ that has $S$ as its inner boundary and extends to
$r=\infty$. Let  $\mathscr{R}_{\text{\em int}}$ be the connected
component of $\{ q \in M: f(q)<0\}$ that has $S$ as its outer
boundary.

Consider first the case of a future endless affine null geodesic
$\gamma (\lambda )$.  The null geodesic equations for the metric
(\protect\ref{metricform}) give
\begin{equation}\label{gennullgeodesic}
\frac{d^2r}{d\lambda^2} = rf(r) \left\{ \left(\frac{d \theta}{d
\lambda}\right)^2 + \sin^2\theta\left(\frac{d \phi}{d
\lambda}\right)^2\right\} -\frac{\mu '(r)}{2}\left(\frac{dr}{d
\lambda}\right)^2 \ .
\end{equation}
At any point $p \in |\gamma |\cap \mathscr{R}_{\text{\em ext}}$
such that $dr/d\lambda =0$ one has $d^2r/d\lambda^2 >0$. At any
point $p \in |\gamma |\cap \mathscr{R}_{\text{\em int}}$ such
that $dr/d\lambda =0$ one has $d^2r/d\lambda^2 <0$. Thus if
$\gamma$ starts outside $S$ (i.e.\ in $\mathscr{R}_{\text{\em
ext}}$) and is initially directed outwards, in the sense that
$dr/d\lambda$ is initially positive, then $\gamma$ will continue
outwards. If $\gamma$ starts in $\mathscr{R}_{\text{\em int}}$
and is initially directed inwards, in the sense that
$dr/d\lambda$ is initially negative, then $\gamma$ will continue
inwards until it falls either into a singularity or through an
$SO(3)\times {\mathbb{R}}$-invariant photon surface other than $S$.

Consider now a unit speed timelike geodesic $\xi (s)$. The
timelike geodesic equations give
\begin{equation}\label{gentimelikegeodesic}
\frac{d^2r}{ds^2} = - \frac{m(r)}{r^2} -4\pi rp_1(r) + rf(r)
\left\{ \left(\frac{d \theta}{ds}\right)^2 +
\sin^2\theta\left(\frac{d \phi}{ds}\right)^2\right\} - \frac{\mu
'(r)}{2}\left( \frac{dr}{ds} \right)^2 \ .
\end{equation}
For any point $p \in \mathscr{R}_{\text{\em ext}}$ one can
arrange for the first three terms on the right of
(\protect\ref{gentimelikegeodesic}) to cancel and so obtain a
unit speed timelike geodesic $ \xi (s)$ through $p \in {\mathcal
R}_{\text{\em ext}}$ at constant $r$. For $p \in {\mathcal
R}_{\text{\em int}}$ the first three terms on the right of
(\protect\ref{gentimelikegeodesic}) are evidently negative. For
$p \in | \xi |\cap \mathscr{R}_{\text{\em int}}$ such that $dr/ds
= 0$ one has $d^2r/ds^2 <0$. Thus if $\xi$ starts in ${\mathcal
R}_{\text{\em int}}$ and is initially directed inwards, in the
sense that $dr/ds$ is initially negative, then $\xi$ will
continue inwards until it falls either into a singularity or
through an $SO(3)\times {\mathbb{R}}$-invariant photon surface other
than $S$.

\protect\section{Spherical symmetry: Examples}\label{ssex} The
following are some examples of $SO(3)\times {\mathbb{R}}$-invariant and
$SO(3)$-invariant photon surfaces in familiar space-times.

\begin{example}[Schwarzschild space-time]\label{Schwarzschildex}
{\rm The metric of Schwarzschild space-time in single null
(radiation) coordinates has the form \begin{equation}
 {\bf  g} =
-\left(1-\frac{2m}{r}\right) du^2
            + 2 du dr
       + r^2
\left(d\theta^2 + \sin^2\theta d\phi^2 \right) . \label{Schmetric}
\end{equation} In this case equation $(\protect\ref{photonsphere})$ reduces to $r =
3m$. The timelike hypersurface $\{ r=3m\}$ is thus an
$SO(3)\times {\mathbb{R}}$-invariant photon surface, or photon sphere,
as expected.  There are no other
$SO(3)\times {\mathbb{R}}$-invariant timelike photon surfaces.

For a non-zero cosmological constant $\Lambda$, the Schwarzschild
metric (\protect\ref{Schmetric}) generalizes to the
Schwarzschild-de Sitter metric

\begin{equation}
{\bf  g} = -\left(1-\frac{2m}{r}+\frac{\Lambda r^2}{3}\right) du^2
           + 2 du dr
      + r^2 \left(d\theta^2 + \sin^2\theta d\phi^2 \right) \ .
\label{schdesit}
\end{equation}
One finds that equation (\protect\ref{photonsphere}) reduces to
$r=3m$, independent of the value of $\Lambda$.  This is
surprising since Schwarzschild and Schwarzschild-de Sitter
space-times are not conformally related. }
\end{example}
\begin{example}[Schwarzschild interior solution]
{\rm The Schwarzschild interior solution describes a spherically
symmetric distribution of perfect fluid of radius $R$, bounded
pressure $p$ and constant density  $\rho_0 >0$. The solution is
to have a metric of the form (\protect\ref{metricform}) and is to
be matched at $r=R$ to a Schwarzschild vacuum solution in such a
way that the pressure is a continuous function of $r$. One thus
has $p_1(r)=p_2(r)=:p(r)$ for all $r:0\leq r<\infty$, $\rho
(r)=\rho_0$ for all $r:0\leq r\leq R$, $\rho (r)= 0 $ for all $r:
R <r<\infty$ and  $p(r)=0$  for all $r: R \leq r<\infty$. The
pressure $p(r)$ for $r:0\leq r\leq R$ is to be obtained by an
integration of the Tolman-Oppenheimer-Volkoff equation
(\protect\ref{TOV}) subject to the boundary condition $p(R)=0$.
This yields
\begin{align}
m(r) &= \frac{\mbox{\normalsize $4\pi$}}{\mbox{\normalsize $3$}} \rho_0 r^3 \ ,
& 0\leq r \leq R \ , \label{mconden} \\[2ex]
{\rm e}^{\mu (r)} &=
\frac{\mbox{\normalsize $(3-u(r))^2$}}{\mbox{\normalsize $4u^2(r)$}} \ ,
& 0\leq r \leq R \ , \\[2ex]
p(r) &= \frac{\mbox{\normalsize $u(r)-1$}}{\mbox{\normalsize$ 3-u(r)$}} \ \rho_0 \ ,
& 0\leq r \leq R \ , \label{pconden}\\
\intertext{for}
u(r) &:= \left(\frac{3-8\pi\rho_0 r^2}{3-8\pi\rho_0
R^2}\right)^{\frac{1}{2}} \ ,
& 0\leq r \leq R \ . \label{udef}
\end{align}
The spherically symmetric system described by the Schwarzschild
interior solution can exist in a state of stable equilibrium iff
$m(R)/R < 4/9$ (see in Stephani \protect\cite{Ste82}). This
condition is equivalent to $8 \pi \rho_0 R^2 < 8/3$, which implies
$p(r)\geq0$ $\forall r\geq 0$, and implies that the absence of a
black hole is a general feature of the Schwarzschild interior
solution.

The left side of equation (\protect\ref{fluidphotonsphere}) now
assumes the form
\begin{equation} \label{intSchps}
1-\frac{3m(r)}{r}-4\pi r^2 p(r) = \left\{ \begin{array}{ll}
1-\frac{\mbox{\normalsize $8\pi\rho_0r^2$}}{\mbox{\normalsize
$3-u(r)$}} &: 0< r \leq R \ ,\\[2ex] 1-\frac{\mbox{\normalsize
$3m(R)$}}{\mbox{\normalsize $r$}} & : r \geq R \ . \end{array}
\right. \end{equation} For $m(R)/R <1/3$ one has that
(\protect\ref{intSchps}) is positive for all $r>0$, so there are
no timelike photon spheres. For $m(R)/R=1/3$ there is a single
timelike photon sphere which lies at the boundary $r=R$ of the
matter. For $1/3< m(R)/R<4/9$ there is one timelike photon sphere
outside the matter at $r=3m(R)>R$ and one timelike photon sphere
inside the matter at
\begin{equation}
r= \left( \frac{1-3\pi\rho_0
R^2}{\pi\rho_0(3-8\pi\rho_0R^2)}\right)^{1/2}
  =\frac{2R}{3}
  {\left(
   \frac{   1-\frac{9m\left(R\right)}{4R}  }
        {
        \left(1-\frac{2m\left(R\right)}{R}\right)
\frac{m\left(R\right)}{R}
        }
  \right)}^{1/2}      <R \ .
\end{equation}
For fixed $R$ the radius of the outer photon sphere is a strictly
increasing function of $m(R)/R$ whilst the radius of the inner
photon sphere is a strictly decreasing function of $m(R)/R$.

Thus a Schwarzschild interior solution matched to a Schwarzschild
vacuum exterior solution contains no black hole, and contains one
timelike photon sphere iff $1/3 = m(R)/R$ and two timelike photon
spheres iff $m(R)/R$ lies in the range $1/3 < m(R)/R<4/9$.
However for such values of $m(R)/R$ the space-time is unphysical
in that the pressure at the center is $p(0)\geq
\rho_0/\sqrt{3}>\rho_0/3$. Therefore under the reasonable energy
condition $p(r) \leq \rho_0/3$ $\forall r: 0 \leq r < \infty$
there are no photon spheres in this example.

The energy condition $0\leq p(r)\leq \rho (r)/3$ $\forall r \in
[0,\infty )$ is in fact satisfied iff $m(R)/R$ lies in the range
$0\leq m(R)/R \leq 5/18$. The corresponding space-times (in view
of $5/18 <7/24$) satisfy all of conditions (1) to (4) of Theorem
\protect\ref{nopstheorem}, except the smoothness of $\rho (r)$ at
$r=R$. Thus considering a smoothed family of solutions
approximating the solution above and having it as a strict limit,
but each with smooth $\rho (r)$ at $r=R$, we can apply Theorem
\protect\ref{nopstheorem} to show no photon spheres exist in all
the cases discussed in this section where this energy condition
is satisfied. On the other hand Proposition
\protect\ref{nopsprop} is directly applicable without smoothing,
but only applies to the subset of these cases with $0\leq m(R)/R
< 1/4$.}
\end{example}
\begin{example}[Reissner-Nordstr{\"o}m space-time]\label{rnex}
{\rm The general static spherically symmetric solution to the Einstein-Maxwell equations
is the Reissner-Nordstr{\"o}m solution comprising a metric ${\bf g}$
and an electromagnetic field $F_{ab}$ given by
\begin{align}
{\bf  g} &= -\left(1-\frac{2m}{r}+\frac{e^2}{r^2}\right) du^2
           + 2du dr
      + r^2 \left(d\theta^2 + \sin^2\theta d\phi^2 \right) \label{rn} \\
F_{tr}&=-F_{rt} = \frac{e}{r^2} \ , \quad \text{\em all other
components vanishing,}
\end{align}
where $m$ is the ADM mass and $e$ is the electric charge. We
assume $m>0$. There is an event horizon at
$r=r_+:=m+\sqrt{m^2-e^2}$ and a Cauchy horizon at
$r=r_-:=m-\sqrt{m^2-e^2}$. The event horizon exists for $0\leq
(e/m)^2\leq 1$, and the Cauchy horizon exists for $0<(e/m)^2\leq
1$. For $(e/m)^2=1$ they both lie at $r=m$. There can be no
timelike photon spheres between the event horizon and the Cauchy
horizon because the Killing field ${\bf K}$ is spacelike there.
Outside the event horizon the Killing field ${\bf K}$ is
timelike, so every hypersurface of the form $\{ r=\text{\em
const.}\}$ which lies outside the event horizon and satisfies the
photon sphere equation (\protect\ref{photonsphere}) is
necessarily a timelike photon sphere.

Equation (\protect\ref{photonsphere}) assumes the form
\begin{equation}
r^2 - 3 m r + 2 e^2 =0
\end{equation}
which has solutions $r_{\text{\em ps}}^{\pm}$ given by
\begin{equation}
r_{\text{\em ps}}^{\pm}/m =  \frac{3 {\pm}\sqrt{9-8(e/m)^2}}{2} \ .
\label{rnphotonsphere}
\end{equation}
The hypersurface $S^+ := \{ r=r_{\text{\em ps}}^+\}$ exists for
$0\leq (e/m)^2\leq 9/8$ and lies outside the event horizon and is
therefore a timelike photon sphere. The hypersurface $S^- :=\{
r=r_{\text{\em ps}}^-\}$ exists for $0< (e/m)^2\leq 9/8$ but lies
outside the event horizon only for $1< (e/m)^2 \leq 9/8$, and so
is a timelike photon sphere only then. The hypersurfaces $S^+$
and $S^-$ coincide for $(e/m)^2=9/8$. The Cauchy horizon and event
horizon are always null photon spheres.

For $0\leq (e/m)^2 \leq 1$ the curvature singularity at $r=0$ is
locally naked but hidden behind an event horizon which lies
strictly inside the only timelike photon sphere. For $1<(e/m)^2$
the singularity at $r=0$ is globally naked and is surrounded by
two timelike photon spheres in the case $1<(e/m)^2 < 9/8$, one
timelike photon sphere in the case $(e/m)^2=9/8$ and by no photon
spheres, either timelike or null, in the case $(e/m)^2> 9/8$ (see
Fig.~\protect\ref{rnfigure}).}
\begin{figure}[t]
\begin{center}
\includegraphics[width=4.0in]{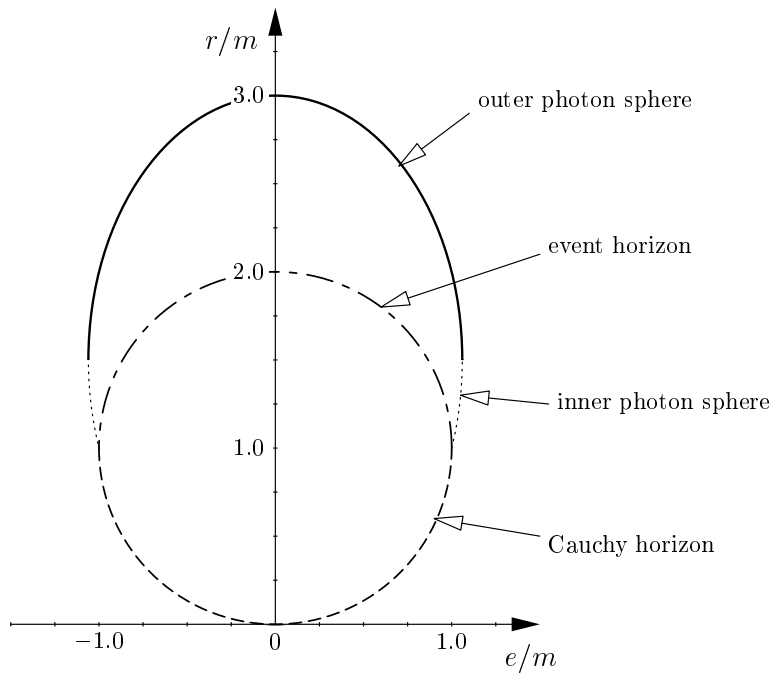}
\end{center}
\caption{The radii of the event horizon, Cauchy
horizon and photon spheres for Reissner-Nordstr{\"o}m space-time.}
\label{rnfigure}
\end{figure}
\end{example}
\begin{example}[Janis-Newman-Winicour space-time]
{\rm The most general static spherically symmetric solution to
the Einstein massless scalar field equations for a scalar field
$\Phi$ satisfying $\square \Phi =0$ was obtained by Janis, Newman
and Winicour \protect\cite{JNW}. The Ricci tensor has the form
$R_{ab}=8\pi \Phi_{;a}\Phi_{;b}$. The solution is known
\protect\cite{Vir97} to be expressible in the form
\begin{align} {\bf g} &= -\left(1-\frac{b}{r}\right)^{\nu}
dt^2
      +\left(1-\frac{b}{r}\right)^{-\nu} dr^2
      + \left(1-\frac{b}{r}\right)^{1-\nu} r^2 \left(d\theta^2
                   +\sin^2\theta \  d\phi^2\right) \\
\Phi &= \frac{q}{b \sqrt{4\pi}} \ln \left(1-\frac{b}{r}\right)
\label{jnwle}
\end{align}
for $r: b<r<\infty$ where the constants $b$, $\nu$ are related to
the ADM mass $m$ and scalar charge $q$ by
\begin{equation}
\nu = \frac{2m}{b}, \quad b = 2 \sqrt{m^2+q^2}\ .
\end{equation}
We assume $b>0$.  There is a curvature singularity at $r=b$. In
order to obtain $m \geq 0$ one must assume $0\leq \nu \leq 1$.
For $q=0$ the solution reduces to the Schwarzschild solution.

Since all hypersurfaces of the form $\{ r=\text{\em const.}\}$
are timelike, one has from the photon sphere equation
$(\protect\ref{photonsphere})$ that the only timelike photon
sphere is at
\begin{equation} r= \frac{b \left(2\nu +1\right)}{2} \ ,
\label{jnwps}
\end{equation}
which exists only for $\nu : \frac{1}{2} < \nu \leq 1$, i.e.\ for
$0\leq q^2<3m^2$. For $\frac{1}{2} < \nu \leq 1$ it is known
\protect\cite{Viretal98} that a photon coming from infinity is
deflected through an unboundedly large angle, i.e.\ the photon
passes increasingly many times around the singularity as the
closest distance of approach tends to the right side of
$(\protect\ref{jnwps})$. } \end{example}
\begin{example}[Charged dilaton space-time]
{\rm A static spherically symmetric space-time with a charged dilaton
field was obtained by Horne \& Horowitz \protect\cite{HH92}.
It comprises a metric ${\bf g}$, a dilaton field
$\Phi$ and an electromagnetic field $F_{ab}$ given by
\begin{align}
{\bf g} &= -\left(1-\frac{r_{+}}{r}\right)
\left(1-\frac{r_{-}}{r}\right)^{\omega} dt^2
    + \left(1-\frac{r_{+}}{r}\right)^{-1}
       \left(1-\frac{r_{-}}{r}\right)^{-\omega} dr^2 \nonumber \\
  &\hspace*{18em} +  \left(1-\frac{r_{-}}{r}\right)^{1-\omega} r^2 d\Omega^2
\label{cdmetric}\\ {\rm e}^{2\Phi} &=
\left(1-\frac{r_{-}}{r}\right)^{(1-\omega)/\beta}  \\
F_{tr}&=-F_{rt} = \frac{e}{r^2} \ , \quad \text{\em  all other
components vanishing,} \label{Fcd}
\end{align}
where $r_{+}$ and $r_{-}$ are related to the ADM mass $m$ and
electric charge $e$ by
\begin{align}
r_{+} + \omega r_{-}&= 2m \label{rprm1} \\ r_{+} r_{-} &= e^2
(1+\beta^2)  \label{rprm2}
\end{align}
and $\beta$ is a free parameter which controls the coupling
strength between the dilaton and Maxwell fields, with $\omega$
defined in terms of $\beta^2$ by
\begin{equation}
\omega  :=  \frac{1-{\beta}^2}{1+{\beta}^2} \ . \label{omega}
\end{equation}
We assume $m>0$.

For $\beta=0$ the solution reduces to the Reissner-Nordstr{\"o}m
solution considered in Example \protect\ref{rnex}. For $\beta =0$
and $e=0$ the solution reduces to the Schwarzschild solution.  The
solution also reduces to the Schwarzschild solution for $e=r_-=0$
and arbitrary $\beta$. Here we shall consider the case $\beta^2 =
1$. In this case one has $(r_+,r_-)=(2m,e^2/m)$. There is an event
horizon at $r = r_+=2m$ and a curvature singularity at $r= r_- =
e^2/m$. For $0\leq (e/m)^2<2$ the singularity at $r= r_-$ lies
inside a black hole whilst for $(e/m)^2 > 2$ it is globally naked.

For $\beta^2=1$ the photon sphere equation
(\protect\ref{photonsphere}) reduces to
\begin{equation}
r_{\text{\em ps}}^{\pm}/m = \frac{(6+(e/m)^2){\pm}\sqrt{36+(e/m)^4 -
20(e/m)^2}}{4}  \ .
\end{equation}
For $0\leq (e/m)^2<2$ one has $r_{\text{\em ps}}^- \leq r_- < r_+
< r_{\text{\em ps}}^+$ so there is a single timelike photon
sphere. For $(e/m)^2=2$ one has $r_{\text{\em ps}}^- = r_- = r_+ =
r_{\text{\em ps}}^+$ so there are no timelike photon spheres. For
$2<(e/m)^2<18$ both $r_{\text{\em ps}}^+$ and $r_{\text{\em
ps}}^-$ are complex so there are no timelike photon spheres. For
$(e/m)^2\geq 18$ one has $r_{\text{\em ps}}^- \leq r_{\text{\em
ps}}^+ < r_-$ so there are again no timelike photon spheres. Thus
in the black hole case $0\leq (e/m)^2<2$ there is a single
timelike photon sphere, whilst in the naked singularity case
$(e/m)^2 > 2$ there are no timelike photon spheres.  (See
Fig.~\protect\ref{ghsfig}.)}
\begin{figure}[t]
\begin{center}
\includegraphics[width=4.0in]{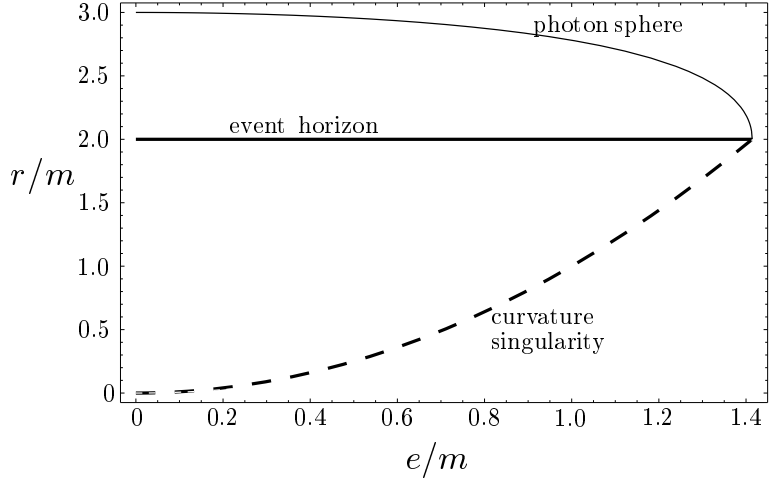}
\end{center}
\caption{The radii of the photon sphere, event horizon and
curvature singularity are plotted against the electric charge
for the charged dilaton solution.}
\label{ghsfig}
\end{figure}
\end{example}

\begin{example}[Vaidya null dust collapse] \label{Vaidyaex}
{\rm The Vaidya null dust collapse model is a non-static,
spherically symmetric space-time with a metric which, in terms of
single null (radiation) coordinates $(u,r,\theta ,\phi )$,
assumes the form
\begin{equation}
{\bf g} = -\left(1-\frac{2m(u)}{r}\right)du^2 +2dudr +
r^2(d\theta^2 +\sin^2 \theta d\phi^2 ) \label{Vmetric}
\end{equation}
where $m(u)$ is a freely specifiable function of $u$.  Setting
$x^0:=u$, $x^1:=r$ in (\protect\ref{x12x0}) one obtains
\begin{align}
\frac{d^2r}{du^2} &= \frac{1}{r}\left(
g_{uu}+\frac{dr}{du}\right) \left( g_{uu}+2\frac{dr}{du}\right)
-\smallfrac{3}{2}\frac{dr}{du}\partial_rg_{uu}
-\smallfrac{1}{2}\left( g_{uu}\partial_rg_{uu} +\partial_ug_{uu}\right)\\
&=\frac{1}{r}\left\{ \left( 1-\frac{3m(u)}{r}\right) \left( 1
-\frac{2m(u)}{r}-3\frac{dr}{du}\right) - \frac{dm(u)}{du}
+2\left( \frac{dr}{du}\right)^2 \right\} \ . \label{r2uV}
\end{align}
This is the evolution equation for a spherically symmetric photon
surface of the Vaidya collapse metric (\protect\ref{Vmetric}).

Consider the special case
\begin{equation}
m(u) = \left\{ \begin{array}{ll} 0&: -\infty < u <0 \\[1ex]
\lambda u & :0\leq u \leq u_1 \\[1ex] m_1:= \lambda u_1 &
:u_1<u<\infty
\end{array} \right.
\end{equation}
for given constants $\lambda >0$, $u_1>0$.  For $u<0$ the
space-time is locally Minkowskian, for $u:0\leq u \leq u_1$ there
is inward falling null dust, and for $u>u_1$ the space-time is
locally isometric to Schwarzschild space-time with ADM mass
$m_1>0$.  It is well-known there is a curvature singularity at
$r=0$ and that for $\lambda : 0<\lambda \leq \frac{1}{16}$ the
part of this singularity at $u=0$ is locally naked.

\begin{figure}[t]
\begin{center}
\includegraphics[width=4.0in]{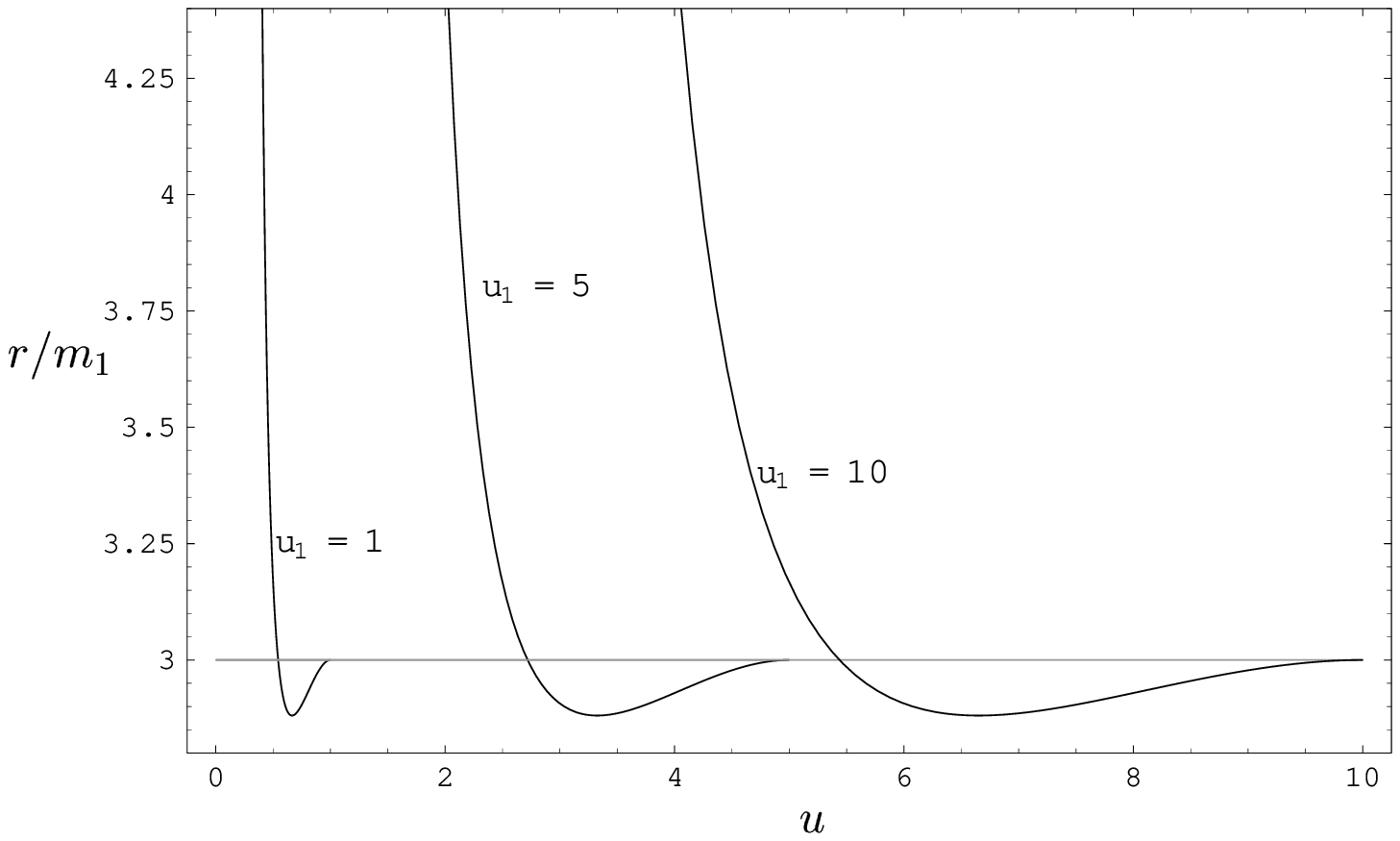}\\[3ex] \includegraphics[width=4.0in]{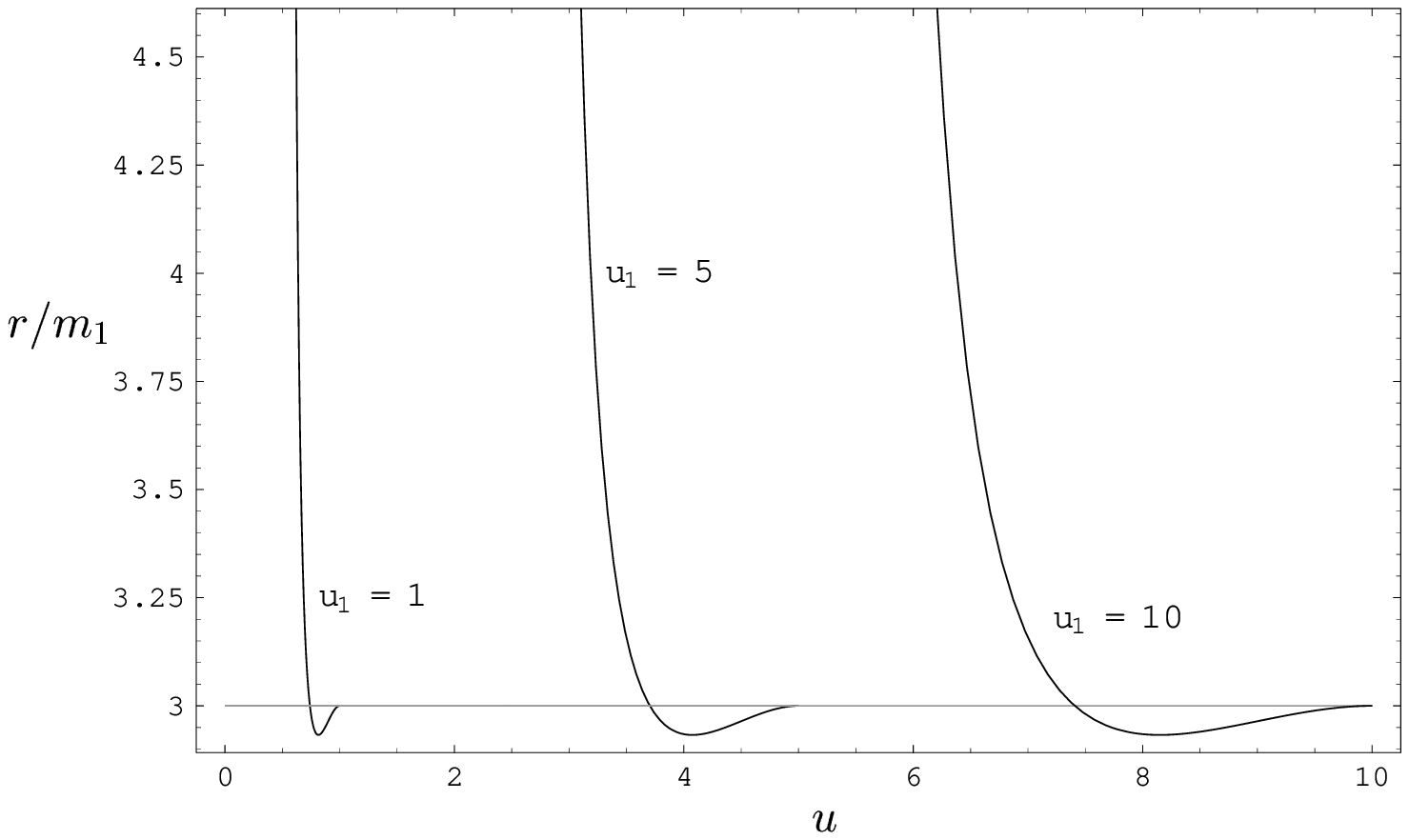}
\end{center}
\caption{The backwards evolution of the $r=3m_1$ Schwarzschild
photon surface through collapsing Vaidya null dust.  The
space-time coordinates of the evolved surface are plotted for
$m(u)=u /16$ (above) and $m(u)=u/32$ (below) for various values of
the null time $u_1>0$ of the junction between the null dust and
Schwarzschild regions.  In each case the evolved photon surface
fails to intersect the Minkowskian region $u<0$.}
\label{vaidyagraphs}
\end{figure}
\begin{figure}[t]
\begin{center}
\includegraphics[width=4.0in]{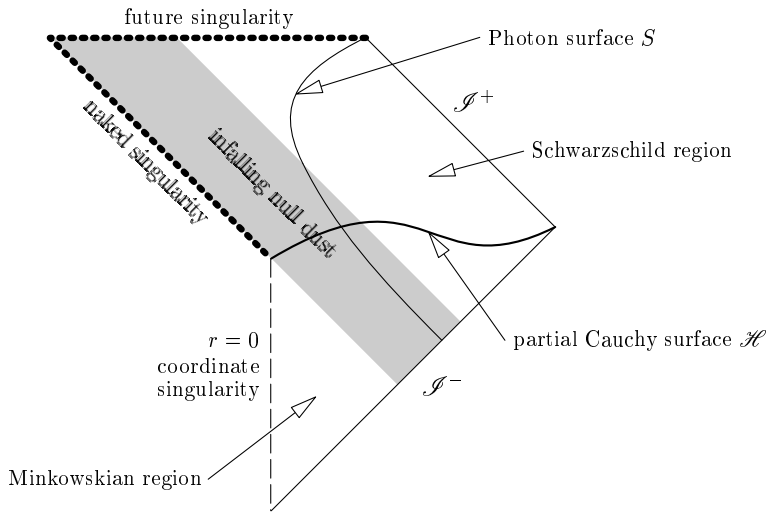}
\end{center}
\caption{The conformal diagram of the Vaidya null dust collapse
model showing the photon surface $S$ arising from the backwards
evolution of the Schwarzschild photon sphere. Any partial Cauchy
surface $\mathscr{H}$ which extends to spatial infinity is cut
into two components by $S$.} \label{pendiag}
\end{figure}

Fix $\lambda : 0<\lambda \leq \frac{1}{16}$. For $u>u_1$ equation
(\protect\ref{r2uV}) gives, as expected, that there is a  photon
surface at $r=3m_1$.  We seek to evolve this photon surface
backwards in time, though the in-falling null dust, to obtain a
maximally extended photon surface $S$. The boundary conditions are
\begin{equation} \left. \begin{array}{c} r=3m_1\\[1ex]
\frac{\mbox{\normalsize $dr$}}{\mbox{\normalsize $du$}}=0
\end{array} \right\} \quad \text{at $u = u_1$} \ .
\end{equation}
The results are shown in Fig.~\protect\ref{vaidyagraphs} for
$\lambda =1/16$ and $\lambda = 1/32$ for selected values of
$u_1$. One sees that in all cases $S$ tends in the past direction
to a null hypersurface of the form $\{ u=\text{\em const.}\}$.
The conformal diagram must therefore be of the form sketched in
Fig.~\protect\ref{pendiag}.

It is evident from Fig.~\protect\ref{pendiag} that the naked
central singularity is enclosed within the photon surface $S$ in
the sense that any partial Cauchy surface $\mathscr{H}$ extending
to spatial infinity must intersect $S$  in a 2-sphere. The
physical significance of this may warrant further investigation.
}\end{example}
\protect\section{Concluding remarks} The definition of a photon
surface given in Section \protect\ref{photonsurfacesection} is
valid in an arbitrary space-time.  However the result that a
photon surface must have a second fundamental form which is pure
trace indicates that a space-time must be specialized in some
respect if it is to contain any photon surfaces.  For spherically
symmetric space-times there are always photon surfaces that
respect the spherical symmetry. For space-times that are not
spherically symmetric,  the definitions of a photon surface and
$G$-invariant photon surface may seem too restrictive. The
problem is that, in general, one may not have orbiting null
geodesics at a fixed radius.  In Kerr space-time for example,
although there are orbiting null geodesics in the equatorial
plane, those null geodesics which move in the direction of
rotation do so at a different radius than those which move in the
opposite direction.  But it seems implausible that a concept
which is physically important in the case of exact spherical
symmetry should become invalid when even a small amount of
angular momentum is introduced.  A non-trivial generalization of
the concepts of photon surface and $G$-invariant photon surface,
at least to axially symmetric space-times, is thus required.

\protect \section*{Acknowledgements}

This research was supported by the NFR of Sweden and NRF of South
Africa.
\bibliographystyle{plain}


\end{document}